\documentclass[11pt,letterpaper]{article}
\usepackage[T1]{fontenc}
\usepackage[utf8]{inputenc}
\usepackage{epic,eepic,amsmath,latexsym,fullpage,amssymb,color,amsthm}
\usepackage{ifthen,graphics,epsfig}
\usepackage{float}
\usepackage{xspace}
\usepackage[english]{babel}
\usepackage{times}
\usepackage{ifpdf}

\newlength{\mymargin}
\usepackage[lmargin=\mymargin,rmargin=\mymargin,tmargin=\mymargin,bmargin=\mymargin]{geometry}

\newlength {\squarewidth}

\newtheorem{theorem}{Theorem}
\newtheorem{lemma}{Lemma}

\newlength {\afterproof}
\newcommand{\toto}{xxx}
\newenvironment{proofT}{\noindent{\bf Proof }}
{\hspace*{\fill}$\Box_{\mathit{Theorem}~\ref{\toto}}$\par\vspace{\afterproof}}
\newenvironment{proofL}{\noindent{\bf Proof }}
{\hspace*{\fill}$\Box_{\mathit{Lemma}~\ref{\toto}}$\par\vspace{\afterproof}}

\newenvironment{lemma-repeat}[1]{\begin{trivlist}
\item[\hspace{\labelsep}{\bf\noindent Lemma~\ref{#1} }]}{\end{trivlist}}

\newenvironment{theorem-repeat}[1]{\begin{trivlist}
\item[\hspace{\labelsep}{\bf\noindent Theorem~\ref{#1} }]}{\end{trivlist}}

\newcommand{\Xomit}[1]{}
\newcounter{linecounter}
\newcommand{\linenumbering}{\ifthenelse{\value{linecounter}<10}
{(0\arabic{linecounter})}{(\arabic{linecounter})}}
\renewcommand{\line}[1]{\refstepcounter{linecounter}\label{#1}\linenumbering}
\newcommand{\resetline}{\setcounter{linecounter}{0}}
\renewcommand{\thelinecounter}{\ifnum \value{linecounter} >
9\else 0\fi \arabic{linecounter}}
\newenvironment{mycenter}{\begin{list}{}{\topsep0pt \partopsep0pt \parskip0pt \leftmargin0pt}\centering\item\relax}{\end{list}}
\newsavebox{\mybox}
\newenvironment{myalgo}[1][aaaaa\=aa\=aa\=aa\=aa\=\kill]{
\begin{lrbox}{\mybox}
\footnotesize

\renewcommand{\baselinestretch}{2.5}
\begin{minipage}[t]{0pt}
\begin{tabbing}#1
}{\end{tabbing}
\end{minipage}\normalsize
\renewcommand{\baselinestretch}{\oldbaselinestretch}
\end{lrbox}
\begin{mycenter}
\fbox{\usebox{\mybox}}
\end{mycenter}
}

\newcommand{\CM}{{\cal CAMP}_{n,t}}
\newcommand{\CMprim}{{\cal CAMP}_{n',t}}
\newcommand{\BM}{{\cal BAMP}_{n,t}}

\newcommand{\SA}{\mathit{SA}}

\begin{document}

\title{\bf From Byzantine Failures to Crash Failures in\\ 
             Message-Passing Systems: 
            a BG Simulation-based approach}
\author{Damien Imbs$^{\dag}$,     Michel Raynal$^{\star,\ddag}$, Julien Stainer$^{\bullet}$\\~\\
$^{\star}$ Institut Universitaire de France\\
$^{\dag}$ Department of Mathematics, University of Bremen, Germany\\
$^{\ddag}$ IRISA, Université de Rennes, 35042 Rennes, France \\
$^{\bullet}$ École Polytechnique Fédérale de Lausanne, Switzerland\\
{\footnotesize {\tt
imbs@uni-bremen.de~~~raynal@irisa.fr~~~julien.stainer@epfl.ch
}}
}

\date{}
\maketitle

\begin{abstract}
The BG-simulation  is a powerful reduction
algorithm designed for asynchronous read/write crash-prone systems.
It allows a  set of $(t+1)$ asynchronous
sequential processes to wait-free simulate (i.e., despite the crash of up
to $t$ of them) an arbitrary number $n$ of processes under the assumption
that at most $t$ of them may crash.  
The BG simulation shows that, in
read/write  systems, the  crucial parameter is not the number $n$ of processes, 
but the upper bound $t$ on the number of process crashes.

The paper extends the concept of BG simulation to asynchronous message-passing 
systems prone to Byzantine failures.
Byzantine failures are the most general type of failure: a faulty process 
can exhibit any arbitrary behavior. Because of this, they are also the most 
difficult to analyze and to handle algorithmically. 
The main contribution of the paper 
is a signature-free reduction of Byzantine failures to crash failures.
Assuming $t<\min(n',n/3)$,  the paper presents an algorithm that simulates 
a system of
$n'$ processes where up to $t$ may crash, on top of a basic system
of $n$ processes where up to $t$ may be  Byzantine. 
While topological techniques have been used to relate the computability of
Byzantine failure-prone systems to that of crash failure-prone ones, 
this simulation is the first, to our knowledge, that establishes this relation 
directly, in an algorithmic way.

In addition to extending the basic BG simulation to  message-passing systems
and failures more severe than process crashes, being modular and direct,
this simulation provides us with a deeper insight in  the nature and
understanding of crash and Byzantine failures in the context of
asynchronous message-passing systems. Moreover, it 
also allows crash-tolerant algorithms, designed for   
asynchronous read/write systems,  to be executed on top of asynchronous
message-passing systems  prone to  Byzantine failures.\\

\end{abstract}

\thispagestyle{empty}
\newpage
\setcounter{page}{1}

\section{Introduction}
\paragraph{What is the Borowsky-Gafni (BG) simulation and why is it important?}
Considering an asynchronous system where processes can crash, the $(n,k)$-set
agreement problem is a basic distributed decision task defined as
follows~\cite{C93}.
Each of  the $n$ processes  proposes a value,  and every process  that does
not crash has  to decide a value (termination), such  that a decided value
is a proposed value (validity) and at most $k$ different values are decided
(agreement). The consensus problem corresponds to the particular case $k=1$.

The $(n,k)$-set agreement is fundamental because it captures the essence 
of fault-tolerant distributed computability issues. 
A central  question related to asynchronous distributed computability is
the following: 
{\sl ``Can we use a  solution to the $(n,k)$-set
agreement problem as a subroutine to solve the  $(n',k')$-set
agreement problem, when at most $t<\min(n,n')$ processes may crash?''}
(``Is $(n',k')$-set agreement  reducible to $(n,k)$-set agreement?''.)
The BG simulation (initially sketched in~\cite{BG93}
and then formalized  in a journal version~\cite{BGLR01},
 where, in addition, a formal definition of ``reducibility'' is given) 
answers this  fundamental question.
It states that the answer is ``yes'' if $k' \geq k$ and  ``no'' if
$k' \leq t <k$. As we can see, the answer ``yes'' does not depend on
the number of processes.

To this end, the algorithm described in~\cite{BGLR01} allows $(t+1)$
processes  to simulate a large number $n'$ of asynchronous processes that
communicate through  read/write registers, and collectively solve a decision
task, in the presence of at most $t$ crashes.
Each of the  $(t+1)$ simulator processes simulates all the $n'$ processes.
These $(t+1)$ simulator processes cooperate through underlying objects
that allow them to agree on a single output for each  of the non-deterministic
statements  issued by every simulated process. (These underlying objects,
called safe agreement objects, can be built of top of read/write atomic
registers.)

Let  BG(RW,C) denote the  basic BG simulation algorithm~\cite{BGLR01} (RW
stands for ``read/write communication'', and C stands for ``crash failures'').
BG(RW,C) is ``symmetric'' in  the sense that  each of the  $n'$
processes is  simulated by every simulator, and  the $(t+1)$ simulators
are ``equal'' with respect  to each  simulated process, namely,
(1)  every simulator fairly  simulates all the processes,
and (2) the  crash  of  a simulator  entails the crash  of  at most  one
simulated process.  This symmetry allows  BG(RW,C) to be
suited to  colorless tasks (i.e., distributed computing problems where the
value decided by a process can be decided by any process~\cite{HR97}).
BG(RW,C) has then been extended to colored tasks
(i.e.,  tasks such as renaming~\cite{ABDPR90}, where a process cannot
systematically borrow its output from another process).
Extended BG simulation is addressed in~\cite{G09,IR09}.
Algorithmic pedagogical
presentations of the BG simulation can be found in~\cite{HRR13,IR09}.
A topological view on distributed computability issues in Byzantine
asynchronous message-passing systems has been  recently
presented in~\cite{HKR14,MTH14}.  A pedagogical topology-based 
presentation of the BG-simulation is given in chapter 7 of~\cite{HKR14}.

\paragraph{What is learned from the BG simulation}
The  important  lesson  learned  from the BG simulation is  that, in a
failure-prone  context,  what is important  is  not the number of processes
but  the maximal  number of   possible failures  and the  actual  number of
values that are proposed to a  decision task.
An interesting  consequence  of  the  BG simulation  (among  several
of its  applications  \cite{BGLR01})  is  the  proof  that  there is no
$t$-resilient $(n,k)$-set agreement algorithm for  $t  \geq k$.
This is obtained as follows.  As (1)
the BG simulation allows reducing the $(k+1,k)$-set agreement problem to the
$(n,k)$-set agreement problem in  a system with up to  $k$ failures,
and (2) the $(k+1,k)$-set agreement problem is known to be impossible
in presence of $k$ failures \cite{BG93,HS99,SZ00}, it follows that there is
no   $k$-resilient $(n,k)$-set agreement algorithm.

\paragraph{Content of the paper: on the BG-simulation side}
As already indicated, the BG simulation has been explored in asynchronous
systems where processes
(1) communicate through atomic read/write registers~\cite{L86}, and
(2) may commit  only  crash failures. This paper extends it in two directions.
The first is the communication model, namely, it considers
that processes cooperate by sending and receiving messages
via asynchronous reliable channels.
The second dimension is related to the type of failures; more precisely, it
considers two  types of failures: process crash failures, and
the more severe  process Byzantine failures.
The  paper presents the following contributions.

A first is an algorithm,  denoted BG(MP,C), which simulates the execution 
of a colorless task running in an asynchronous message-passing system of $n'$
processes, where up to $t$ may crash, on top of an  asynchronous
message-passing system of $n$  processes where up to $t$ may crash.
This simulation  requires $t<n/2$ (which is a necessary and sufficient
condition to  simulate read/write registers in asynchronous message-passing
systems of $n$ processes~\cite{ABD95}). While the number of simulated
processes $n'$ can be any integer, for the simulation to be non-trivial
we consider that $t<n'$.

A second contribution is an
algorithm, denoted BG(MP,B), which simulates the execution of a
colorless task running in an asynchronous message-passing system of $n'$
processes, where up to $t$ may crash, on top of an asynchronous message-passing 
system of $n$ processes where up to $t$ may be Byzantine~\cite{LSP82}.
This simulation requires $t<n/3$ (according to the task 
which is simulated, additional constraint on $t$ may be needed, 
see~\cite{HKR14}; see also Section~\ref{sec:conclusion}). 
As in the case of  BG(MP,C), and for the same reason, we consider that $t<n'$.
This algorithm has two noteworthy features: it is the first BG simulation
algorithm that considers Byzantine failures, and it allows to run a
crash-tolerant algorithm solving a colorless task on top of an asynchronous
system prone to Byzantine failures.
Both the algorithms BG(MP,C) and  BG(MP,B) are {\it genuine} in the 
sense they do not rely on the simulation of an underlying shared memory.

While the  full-information algorithm presented
in~\cite{MTH14} can be used to decide when there is a simulation
between two models, the present paper is the first (to our knowledge)
that allows the direct execution in the presence of Byzantine failures
of any crash-tolerant algorithm that solves a colorless task.
BG(MP,B) provides an algorithmic approach which complements the
topology-based simulation framework of~\cite{MTH14}, and may also be
of practical interest.  It has the interesting property that the
simulation of a message only requires a polynomial number of messages
in the base system, and the increase in size of these messages, when
compared to the size of the simulated message, is also polynomial.
Additionally, differently from early works on Byzantine failures
like~\cite{GMW87}, it does not use any cryptography-based mechanism.

\paragraph{Content of the paper: on the safe agreement objects side}
The core of the previous algorithms  lies in 
  new  underlying safe agreement objects, which allow the
$n$ simulators to agree on the next operation executed by each of the $n'$
simulated  processes. Such a  safe agreement object ensures that all the
simulators produce the very same simulation. At the operational level,
a safe agreement object
provides processes with two operations, denoted ${\sf propose}()$
and ${\sf decide}()$, which  are invoked in this order by each correct process.
The termination property associated with  a safe agreement object $\SA$ is
the following: if no simulator commits a failure while executing
$\SA.{\sf propose}()$, then any invocation of $\SA.{\sf decide}()$ 
by a non-faulty simulator terminates. Moreover, no two correct processes 
decide differently. 

On the algorithmic side, a novelty of the paper lies in the algorithms
implementing these new safe agreement objects. Differently from their
read/write memory counterparts, they are not based on underlying
snapshot objects~\cite{AADGMS93}.  They instead rely heavily on
message communication patterns inspired from the reliable broadcast
algorithms  introduced in~\cite{B87}.

A last and noteworthy contribution of the paper lies in the second algorithm
(which implements safe agreement in a  Byzantine message-passing system).
This object is the core of a simulation when one wants to execute     
asynchronous read/write crash-tolerant algorithms  on top of asynchronous
message-passing systems prone to Byzantine failures.

\paragraph{Existing simulations considering Byzantine failures}
Simulations of crash failures in a Byzantine system have been
addressed in the context of synchronous systems~\cite{BN91,NT90,ST87}.
The only articles we are aware of concerning such a simulation in
asynchronous systems are~\cite{C88,HKR14,HDR07}.  As noticed
in~\cite{AW04}, \cite{C88} considers a restricted class of round-based
deterministic algorithms.  The
simulation presented in~\cite{HKR14} executes a full-information
asynchronous crash-tolerant algorithm in an asynchronous Byzantine
failure-prone system. The article~\cite{HDR07} considers an
agent/host model and focuses mainly on reliable broadcast.

\paragraph{Roadmap}
The paper is composed of~\ref{sec:conclusion} sections.
Section~\ref{sec:model} presents both the crash-prone and the Byzantine
asynchronous message-passing models, and the notion of a task.
Section~\ref{sec:simu-structure} presents the structure of the simulation
algorithms. Section~\ref{sec:BG-crash-model} presents the  simulation
algorithm BG(MP,C), while Section~\ref{sec:BG-byzantine-model}
presents the simulation algorithm  BG(MP,B).
Finally, Section~\ref{sec:conclusion}  addresses 
the computability implications of the  Byzantine-tolerant 
simulation and its underlying safe agreement object.

\section{Computation Models and Tasks}
\label{sec:model}

\subsection{Computation models}

\paragraph{Computing entities}
The system is made up of a set $\Pi$ of $n$ sequential processes,
denoted $p_1$, $p_2$, ..., $p_n$. These processes are asynchronous
in the sense that each process  progresses at  its own speed,
which can  be arbitrary and remains always unknown to the other processes.

During an execution, processes may deviate from their specification.
In that case, the corresponding processes are said to be {\it faulty}.
A process that does not deviate from its specification is {\it correct}
(or {\it non-faulty}).
The model parameter $t$ denotes the maximal number of processes that can
be faulty in a given execution. Two failure types are considered below.

\paragraph{Communication model}
The processes cooperate by sending and receiving messages through
bi-directional channels. The communication network is a complete network,
which means that each process $p_i$  can directly send a message to any
process $p_j$ (including itself).
Each channel is reliable (no loss, corruption, or creation of messages),
not necessarily first-in/first-out, and asynchronous (while the transit
time of each message is finite,  there is no upper bound on message
transit times).

The macro-operation ``${\sf broadcast}$ {\sc type}$(m)$'', where {\sc type} is
a message type and $m$ is its content,  is a shortcut for the following
statement: ``${\sf send}$ {\sc type}$(m)$ to each process (including itself)''.

\paragraph{The process crash failure model}
In the crash failure model, a process may prematurely stop its execution.
A process executes correctly its algorithm until it possibly crashes.
Once crashed, a process remains crashed forever.
It is assumed that at most $t$ processes may crash.
If there is no specific constraint on $t$, the corresponding model
is denoted  $\CM[t<n]$.
When  it is assumed that at most $t<n/2$ processes
may crash,  the corresponding model is denoted $\CM[t<n/2]$.

\paragraph{The Byzantine failure model}
A  Byzantine process is a process that behaves
arbitrarily: it may crash, fail to send or receive messages, send
arbitrary messages, start in an arbitrary state, perform arbitrary state
transitions, etc. Hence, a Byzantine process, which is assumed to send the
same message $m$ to all the processes, can send a message $m_1$ to some
processes, a different message $m_2$  to another subset of processes, and no
message at all to the other processes. Moreover, Byzantine processes can
 collude to ``pollute'' the computation.

It is assumed that Byzantine processes cannot control the network, hence,
when  a process receives a  message, it can unambiguously identify its sender.
As previously, $t$ denotes the upper bound on the number of processes that
may commit Byzantine failures.
If there is no constraint on $t$, the corresponding model
is denoted  $\BM[t<n]$.
When  it is assumed that at most $t<n/3$ processes may be faulty,
the corresponding model is denoted $\BM[t<n/3]$.

\subsection{Decision tasks and algorithms solving a task}

\paragraph{Decision tasks}
The problems we are interested in are called {\it decision tasks}
(the reader interested in a more formal presentation of  decision tasks can
consult the literature, e.g., \cite{BGLR01,HS99}).
In every run, each process proposes a value and the proposed values define
an input vector $I$, where $I[j]$ is the value proposed by process $p_j$.
Let ${\cal I}$ denote the set of  allowed input vectors.
Each process has to decide a value. The decided values define an
output  vector $O$,  such that  $O[j]$ is  the value  decided by  $p_j$.
Let ${\cal O}$ be the  set of  the output vectors.

A decision task is a binary relation $\Delta$ from ${\cal I}$ into ${\cal O}$.
A task is {\it colorless} if,  
when a value  $v$ is proposed by a process
$p_j$ (i.e., $I[j]=v$), then $v$ can be proposed by any number of processes and,
when a value  $v'$ is decided by a process
$p_j$ (i.e., $O[j]=v'$), then $v'$ can be decided by any number of processes.
Consensus, and more generally $k$-set agreement, are colorless tasks.
Otherwise the task is  {\it colored}.  Symmetry breaking and renaming are
colored tasks~\cite{ABDPR90,CRR11,IRR11}.

\paragraph{Algorithm solving a task}
An algorithm solves a task in a $t$-resilient  environment if, given
any $I\in {\cal I}$, (1)  each correct process $p_j$  decides a value $o_j$,
 and (2) there is an output vector $O$ such that
$(I,O)\in \Delta$ where $O$ is defined as follows.
If $p_j$ decides $o_j$, then  ${\cal O}[j]=o_j$.
If $p_j$ does not decide, $O[j]$ is set to any  value $v'$ that preserves
the relation $(I,O)\in \Delta$.

Considering a system of $n$ processes, a
task  is  $t$-resiliently solvable if there is  an
algorithm that solves  it in the presence of at most $t$ faulty processes.
As an example, consensus is  not  $1$-resiliently solvable
in asynchronous crash-prone systems, be the communication medium a set of
read/write registers~\cite{LA87}, or a message-passing  system~\cite{FLP85}.
Differently,  renaming  with $2n-1$ new names is $(n-1)$-resiliently solvable
in asynchronous read/write crash-prone systems~\cite{CR12,HS99}, and
is $t$-resiliently solvable in asynchronous crash-prone message-passing
systems for $t<n/2$~\cite{ABDPR90}.

\section{Structure of the Simulation Algorithms}
\label{sec:simu-structure}

\paragraph{Aim}
Let $A'$ be an algorithm that solves a colorless decision task among $n'$
processes in the  system model  $\CMprim[t<n']$. The aim is to design an algorithm that simulates $A'$ in the
system model  $\CM[t<n/2]$  (resp.,  $\BM[t<n/3]$).
As already indicated, the corresponding simulation algorithm is denoted
BG(MP,C) in the first case, and BG(MP,B) in the second case.

\paragraph{Notation}
A simulated process is denoted $p_j$, where $1 \leq j \leq n'$.
Similarly, a simulator process (``simulator'' in short') is  denoted $q_i$,
where $1 \leq i \leq n$.
The set $\Pi$ denote the set of the simulator indexes,
i.e., $\Pi =\{1,...,n\}$.

The safe agreement objects, build in the simulation and used by the simulators,
are identified with upper case letters, e.g., $\SA$.  The  variables local to
simulator $q_j$ is identified with lower case letters, and the resulting
identifiers are subscripted with $j$.

\paragraph{Behavior of a simulator $q_i$}

Each simulator is given the code of all the simulated processes
$p_1$, ..., $p_{n'}$. It manages $n'$ threads, one associated with each
simulated process, and executes them in a fair way.

The code of a simulated process $p_j$ contains local statements,
send statements, and receive statements. It is assumed that
the behavior of a simulated process $p_j$ is deterministic in the sense
it is entirely defined from its local input (as defined by the task instance),
and the order in which  $p_j$ receives messages.

The simulation has to ensure that (1) all simulators simulate the
same behavior of the set of  simulated processes, and (2) a faulty simulator
entails the failure of at most one simulated process. The way
this is realized depends, of course, on the failure model that is considered.

\section{BG(MP,C): BG in the Crash-prone Asynchronous Message-Passing  Model}
\label{sec:BG-crash-model}

This section presents the algorithm BG(MP,C).
As previously indicated, this algorithm simulates, in the model  $\CM[t<n/2]$,
an algorithm $A'$  solving a task in $\CMprim[t<n']$. It is  made up of two
parts:
an algorithm implementing a safe agreement object, and the simulation itself,
which uses several of these objects to allow the simulators to cooperate.

\subsection{Safe agreement object in $\CM[t<n/2]$: definition}
This object type (or variants of it), briefly sketched in the Introduction,
is at the core of both the BG simulation~\cite{BG93,BGLR01, G09,IR09}, and
the liveness guarantees of concurrent objects~\cite{IR11,IR11-a}.
It is a one-shot object that solves consensus in failure-free scenarios,
and allows processes to agree with  a weak termination guarantee
in the presence of failures.

A safe agreement object provides each simulator $q_i$,
$1\leq i\leq n$,  with two operations
denoted ${\sf propose}()$ and  ${\sf decide}()$, that $q_i$ can invoke
at most once, and in this order;  ${\sf propose}()$ allows $q_i$ to propose
a value, while  ${\sf decide}()$ allows it to decide a value.
Considering the crash failure model, the properties associated with this
object are the following ones.
\begin{itemize}
\vspace{-0.2cm}
\item Validity. A decided value is a proposed value.
\vspace{-0.2cm}
\item Agreement. No two simulators decide distinct values.
\vspace{-0.2cm}
\item Propose-Termination.
An invocation of  ${\sf propose}()$ by a correct simulator terminates.
\vspace{-0.2cm}
\item Decide-Termination.
If no simulator crashes while executing  ${\sf propose}()$, then any
invocation of ${\sf decide}()$ by a correct simulator terminates.
\end{itemize}
It is easy to see that a safe agreement object is a consensus object whose
termination condition is failure-dependent. Algorithms implementing 
safe agreement
objects (or variants of it) can be found in~\cite{BG93,BGLR01,IR11-a}.

\subsection{Safe agreement object in $\CM[t<n/2]$: algorithm}
An algorithm implementing a safe agreement object in $\CM[t<n/2]$ is described
in Figure~\ref{algo-SA-msg-passing-crash}.

\paragraph{Local data structures}
Each simulator $q_i$, $1  \leq  i \leq n$, manages three local
data structures, namely, the arrays $values_i[1..n]$, $my\_view_i[1..n]$,
$all\_views_i[1..n]$, all initialized to  $[\bot,...,\bot]$, where
$\bot$ denotes  a default value that cannot be proposed to the safe agreement
object  by the  simulators.
\begin{itemize}
\vspace{-0.2cm}
\item
The aim of $values_i[x]$ is to contain, as currently known by $q_i$,
the value proposed to the  safe agreement object by the simulator $q_x$.
\vspace{-0.2cm}
\item
The aim of $my\_view_i[x]$ is to contain, as known by $q_i$, the value
proposed to the safe agreement object by the simulator $q_x$, as witnessed
by  strictly more than $\frac{n}{2}$  distinct simulators
(i.e., at least a correct process). 
\vspace{-0.2cm}
\item
The aim of $all\_views_i[x]$ is to contain what to $q_i$'s knows
about the view seen by  $q_x$.
\end{itemize}

\paragraph{Algorithm: the operation ${\sf propose}()$}
The algorithm implementing the operation ${\sf propose}()$ invoked
by a simulator $q_i$ is described at lines~C\ref{SA-C-01}-C\ref{SA-C-14}
(client side)  and lines~C\ref{SA-C-20}-C\ref{SA-C-22} (server side).
This algorithm is made up of three parts.
\\

First part.
A simulator $q_i$ first broadcasts the message {\sc value} $(i,v_i)$, where
$v_i$ is the value it proposes to the safe agreement object
 (line~C\ref{SA-C-01}). Then, it waits until it knows that strictly 
more than $\frac{n}{2}$
simulators know its value (line~C\ref{SA-C-02}).
On its ``server'' side, when $q_i$  receives for the first time
the message {\sc value} $(x,v)$, it first saves $v$ in
$values_i[x]$; then it  forwards  the received message
to cope with the (possible) crash of $q_x$ (this witnesses the fact that
$q_i$ knows the value proposed by $p_x$, line~C\ref{SA-C-20})\footnote{
Let us observe that the lines~C\ref{SA-C-01} and~C\ref{SA-C-20}
implement a reliable broadcast of the  message {\sc value} $(i,v_i)$.
Similarly,  the lines~C\ref{SA-C-12} and~C\ref{SA-C-22}
implement a reliable broadcast of the message {\sc view} $(i,my\_view_i)$.
It is easy to see that the cost of such a reliable broadcast
is $O(n^2)$ messages.}).

\renewcommand{\linenumbering}{\ifthenelse{\value{linecounter}<10}
{(C0\arabic{linecounter})}{(C\arabic{linecounter})}}
\begin{figure}[th!]
\begin{myalgo}
\resetline
{\bf operation} $\mathsf{propose}$ ($v_i$) {\bf is}\\
\line{SA-C-01}
~~\= $\mathsf{broadcast}$ {\sc value} $(i,v_i)$;\\

\line{SA-C-02}
\> {\bf wait} \big({\sc value} $(i,v_i)$    $\mathsf{received}$
             from strictly more than $\frac{n}{2}$ different simulators\big);\\

\line{SA-C-03}
\> {\bf for  each} $x\in [1..n]$  {\bf do}
      $\mathsf{broadcast}$ {\sc read} $(i,x)$  {\bf end for};\\

\line{SA-C-04}
\> {\bf for  each} $x\in [1..n]$  {\bf do} \\

\line{SA-C-05}
\> \>\>  {\bf wait} \big(\={\sc read'answer} $(i,x,\bot)$  $\mathsf{received}$
             from strictly more than $\frac{n}{2}$ different  simulators\\

\line{SA-C-06}
\> \> \>\> $~$ $\vee$ $~$ $\exists~w:$ {\sc value} $(x,w)$ $\mathsf{received}$
             from strictly more than $\frac{n}{2}$ different  simulators\big);\\

\line{SA-C-07}
\> \>\> {\bf if} (predicate of line~C\ref{SA-C-06} satisfied)\\

\line{SA-C-08}
\> \>\> \> {\bf then} \= $my\_view_i[x] \gets w$\\

\line{SA-C-09}
\> \>\> \> {\bf else} \> $my\_view_i[x] \gets \bot$\\

\line{SA-C-10}
\> \>\>\> {\bf  end if}\\

\line{SA-C-11}
\> {\bf end for};\\

\line{SA-C-12}
\>  $\mathsf{broadcast}$ {\sc view} $(i,my\_view_i)$;\\

\line{SA-C-13}
\>  {\bf wait} \big({\sc view} $(i,my\_view_i)$  $\mathsf{received}$
             from strictly more than $\frac{n}{2}$ different simulators\big);\\

\line{SA-C-14}
\> ${\sf return}()$.\\~\\

{\bf operation} $\mathsf{decide}$ () {\bf is}\\
\line{SA-C-15}
\>  {\bf wait}
       \big($\exists$ a non-empty set $\sigma \subseteq \Pi$:\\

\line{SA-C-16}
\> \> ~$\forall~y\in \sigma:~
       \big[ (all\_views_i[y]\neq\bot) ~\wedge~
 \big(\forall~z\in \Pi:~(all\_views_i[y][z]\neq\bot)
                               \Rightarrow(z\in \sigma)\big)\big]$;\\

\line{SA-C-17}
\> {\bf let} $min\_\sigma_i$ {\bf be} the set  $\sigma$ of smallest size; \\

\line{SA-C-18}
\>  {\bf let} $res$ {\bf be} $\min(\{values_i[y] ~:~ y\in min\_\sigma_i\})$;\\

\line{SA-C-19}
\> ${\sf return}(res)$.\\~\\

\%----------------------------------------------------------------------------------------------------------------~\\

{\bf when the message} {\sc value} $(x,v)$
     {\bf is} $\mathsf{received}$   {\bf for the first time}:\\

\>  \%
  ``for the first time'' is with respect to  each pair of values $(x,v)$ \%\\
\line{SA-C-20}
\>  $values_i[x] \leftarrow v;$
           $\mathsf{broadcast}$ {\sc value} $(x,v)$.\\~\\

{\bf when the message} {\sc read} $(j,x)$
     {\bf is} $\mathsf{received}$ {\bf for the first time}:\\

\line{SA-C-21}
\> $\mathsf{send}$ {\sc read'answer} $(j,x,values_i[x])$ $\mathsf{to}~q_j$.
\\~\\

{\bf when the message} {\sc view} $(x,view)$
          {\bf is} $\mathsf{received}$ {\bf for the first time}:\\

\line{SA-C-22}
\>   $all\_views_i[x] \leftarrow view$;
     $\mathsf{broadcast}$ {\sc view} $(x,view)$.

\end{myalgo}
\caption{Safe agreement object in $\CM[t<n/2]$ (code for the simulator $q_i$)}
\label{algo-SA-msg-passing-crash}
\end{figure}

Second part. In this part, $q_i$  builds a local view of the values
proposed by the $n$ simulators. To this end, it first broadcasts  messages
{\sc read} $(i,x)$, $1 \leq x\leq n$, to learn the value proposed by
each simulator $q_x$   (line~C\ref{SA-C-03}). On its server side, when $q_i$
receives such a message, it broadcasts by return its current
knowledge of the value proposed by $q_x$  (line~C\ref{SA-C-21}).

Then, the  simulator $q_i$  builds its local view of the values that have been
proposed. For each simulator $q_x$, $q_i$ waits until it has received
from strictly more than $\frac{n}{2}$ distinct  simulators  the very 
same message, namely, either the message {\sc read'answer} $(i,x,\bot)$, 
or the message {\sc value} $(x,w)$   (lines~C\ref{SA-C-05}-C\ref{SA-C-06}).
In the first case, $q_i$  considers that $q_x$ has not yet proposed a value,
while in the second case it considers that  $q_x$  proposed the value $w$
(let us observe that, while $q_i$ can receive both {\sc read'answer} 
$(i,x,\bot)$ and messages  {\sc value} $(x,w)$, it stops waiting as soon as 
it received strictly more than $\frac{n}{2}$ of one of them) 
(lines~C\ref{SA-C-07}-C\ref{SA-C-10}). 

Third part. Finally, the simulator $q_i$  informs the other simulators
on its local view $my\_view_i[1..n]$. To this  end,  it broadcasts the message
{\sc view} $(i,my\_view_i)$. When it has received the corresponding
``acknowledgments'', $q_i$ returns from its invocation of the operation
${\sf propose}()$ (line~C\ref{SA-C-12}-C\ref{SA-C-14}).
(The behavior of  $q_i$ when it receives a message {\sc view} $(x,view)$
is similar to the one when it receives  a message {\sc value} $(x,v)$.
The only difference is that $values_i[x]$ is now replaced by
$all \_views_i[x]$, line~C\ref{SA-C-22}.)

\paragraph{Algorithm: the operation ${\sf decide}()$}
The algorithm implementing the operation ${\sf decide}()$ is described at
lines~C\ref{SA-C-15}-C\ref{SA-C-19}. It consists in a ``closure'' computation.
A simulator $q_i$ waits until it knows a non-empty set of simulators
$\sigma$  such that (a) it knows their views, and (b) this set is closed
under the relation ``has in its published view the value of'' which means
that the processes whose values appear in a view of a process of $\sigma$
are also in  $\sigma$ (lines~C\ref{SA-C-15}-C\ref{SA-C-16}).

Let us observe that it is possible that, locally,  several sets satisfy this
property. If it is the case, $q_i$ selects the smallest of them.
Let $min\_\sigma_i$ be this set of simulators (lines~C\ref{SA-C-17}).
The  value that is returned by $q_i$ is then the smallest
value among the the values proposed by the simulators in  $min\_\sigma_i$
(lines~C\ref{SA-C-18}-C\ref{SA-C-19}).

\subsection{Safe agreement object in $\CM[t<n/2]$: proof}
This section proves that the algorithm presented in
Figure~\ref{algo-SA-msg-passing-crash} implements a safe agreement object,
i.e., any of its runs in  $\CM[t<n/2]$  satisfies the validity, agreement,
and termination properties, which define it.

\begin{lemma}
\label{lemma:safe-agr-term-propose}
An invocation of  ${\sf propose}()$ by a simulator that does not crash
during this  invocation,  terminates.
\end{lemma}

\begin{proofL}
Let us consider a simulator $q_i$ that does not crash during its invocation of
  ${\sf propose}()$.
Hence, $q_i$ broadcast the message {\sc value} $(i,v_i)$
at line~C\ref{SA-C-01}.
This message is received by strictly more than $\frac{n}{2}$ correct 
simulators, and each
of them  broadcasts this message when it receives it. It follows that
$q_i$ cannot block forever at line~C\ref{SA-C-02}.

Let us now consider the wait statement at lines~C\ref{SA-C-05}-C\ref{SA-C-06}.
There are two cases. Let  {\sc read}~$(i,x)$ be a message
broadcast by  the simulator $q_i$ at line~C\ref{SA-C-03}.
\begin{itemize}
\vspace{-0.2cm}
\item Case 1:
No correct simulator ever receives a message {\sc value}~$(x,-)$.
In this case, each correct simulator  $q_y$ is such that
$values_y[x]$ remains always equal to $\bot$.
It follows  that, when $q_y$  receives the  message {\sc  read}~$(i,x)$,
it sends back to $q_i$ the message {\sc read'answer}~$(i,x,\bot)$
(line~C\ref{SA-C-21}).
As there are strictly more than $\frac{n}{2}$ correct simulators, 
$q_i$ eventually receives the   message  {\sc read'answer}~$(i,x,\bot)$ 
from strictly  more than $\frac{n}{2}$
different  simulators, and the predicate of line~C\ref{SA-C-05}
is then satisfied.
\vspace{-0.2cm}
\item Case 2:
At least one correct simulator $q_y$ receives a message {\sc value}~$(x,v)$.
In  this case,  $q_y$  broadcasts  the message  {\sc  value}~$(x,v)$ when  it
receives it (line~C\ref{SA-C-20}).
It follows from the broadcasts issued at this line  that $q_i$ eventually
receives {\sc value}~$(x,v)$ from strictly more than $\frac{n}{2}$ different 
simulators.
When this occurs, the predicate of line~C\ref{SA-C-06} is satisfied,
and $q_i$ exits the wait statement.
\end{itemize}
As this is true for any message  {\sc read}~$(i,x)$ broadcast by
the simulator $q_i$ at line~C\ref{SA-C-03}, it follows that $q_i$ cannot
remain block forever at lines~C\ref{SA-C-05}-C\ref{SA-C-06}.

Let us finally consider  the lines~C\ref{SA-C-12}-C\ref{SA-C-13}.
As the message  {\sc view} $(i,my\_view_i)$  broadcast by $q_i$
at line~C\ref{SA-C-12} is received by at least all the correct processes,
and  each of them broadcast it when it receives it for the first time,
it follows that $q_i$ receives the message  {\sc view} $(i,my\_view_i)$ from
strictly more than $\frac{n}{2}$ distinct processes, and stops waiting at 
line~C\ref{SA-C-13},
which concludes the proof of the lemma.
\renewcommand{\toto}{lemma:safe-agr-term-propose}
\end{proofL}

\begin{lemma}
\label{lemma:safe-agr-validity}
The value returned by an invocation of ${\sf propose}()$ is a value
that was  proposed by a simulator.
\end{lemma}

\begin{proofL}
Let us observe that (due to its definition) the  set $min\_\sigma$
is non-empty, and (due the first predicate of line~C\ref{SA-C-06})
the simulator indexes $y$ it contains are such that $values_i[y]\neq\bot$.
As, for any of those $y$,  $values_i[y]$ is set to a non-$\bot$ value (only
once) at line~C\ref{SA-C-20}, it follows that $q_i$ received a message
{\sc value} $(y,v_y)$. Hence, the values in the  variables  $values_i[y]$
are values proposed by the corresponding simulators $q_y$.
It follows that the value computed at line~C\ref{SA-C-18} is a value that
was proposed by a simulator, which concludes the proof of the lemma.
\renewcommand{\toto}{lemma:safe-agr-validity}
\end{proofL}

\begin{lemma}
\label{lemma:safe-agr-agreement}
No two invocations of ${\sf decide}()$ return different values.
\end{lemma}

\begin{proofL}
Let us first observe that, due to the reliable broadcast of the messages
{\sc value}~() (lines~C\ref{SA-C-01} and~C\ref{SA-C-20})  and
{\sc view}~() (lines~C\ref{SA-C-12} and~C\ref{SA-C-22}), 
and the fact that a simulator broadcast a single message {\sc value}~$()$, 
we have:
\begin{itemize}
\vspace{-0.2cm}
\item $(values_i[x]\neq \bot) ~\wedge~(values_j[x]\neq \bot)
      ~\Rightarrow~(values_i[x]=values_j[x])$.
\vspace{-0.2cm}
\item $(all\_views_i[x]\neq \bot) ~\wedge~(all\_view_j[x]\neq \bot)
      ~\Rightarrow~(all\_views_i[x]=all\_view_j[x])$.
\end{itemize}

Let us assume, by contradiction, that two simulators $q_i$ and $q_j$
decide different values. This means that the sets  $min\_\sigma_i$
 $min\_\sigma_j$ computed at line~C\ref{SA-C-17} by   $q_i$  and  $q_j$,
respectively,  are different.

Since $min\_\sigma_i$ and $min\_\sigma_j$ are different,
let us consider $z\in min\_\sigma_i\setminus min\_\sigma_j$
(if $min\_\sigma_i \subsetneq min\_\sigma_j$,  swap $i$ and $j$).
According to the closure predicate used at line~C\ref{SA-C-16},
as $z\notin min\_\sigma_j$,
we have  $\forall y\in min\_\sigma_j~:~all\_views_j[y][z]=\bot$.
It follows that any simulator $q_y$ such that $y\in min\_\sigma_j$ does not
fulfill the condition of line~C\ref{SA-C-07} for  $x=z$.
Consequently, $q_y$ received at line~C\ref{SA-C-05} a message
{\sc read'answer}($y,z,\bot$)
from a set of simulators $Q_{y,r(z)}$ of size strictly greater
than $\frac{n}{2}$.
Consequently when $q_y$ executed line~C\ref{SA-C-03} for $x=z$, all the
simulators $q_k$ of $Q_{y,r(z)}$ verified $values_k[z]=\bot$.

When the simulator $q_z$ stops waiting at line~C\ref{SA-C-02}, it received
messages {\sc value}($z$,$v_z$)
(where $v_z$ is the value sent by $q_z$ at line~C\ref{SA-C-01})
from a set $Q_{z,w}$ of strictly more than $\frac{n}{2}$ simulators.
It follows that $Q_{y,r(z)}\cap Q_{z,w}\neq\emptyset$, consequently
there is a  simulator $q_k$ that sent
a message {\sc read'answer}($y,z,\bot$) to $q_y$ and a message
{\sc value}($z$,$v_z$) to $q_z$.
Since $value_k[z]$ is never reset to $\bot$ after being assigned,
the simulator $q_y$ necessarily executed line~C\ref{SA-C-03} for $x=z$
strictly before $q_z$ stops waiting at line~C\ref{SA-C-02}. Consequently
$q_y$ stopped waiting at line~C\ref{SA-C-02} before $q_z$
executes line~C\ref{SA-C-03} for $x=y$. It does so after receiving
messages {\sc value}($y$,$v_y$)
(where $v_y$ is the value sent by $q_y$ at line~C\ref{SA-C-01})
from a set $Q_{y,w}$ of strictly more than $\frac{n}{2}$ simulators
$q_k$, and each of these simulators then verifies $values_k=v_y$.
These simulators do not send {\sc read'answer}($z,y,\bot$) messages when
they receive the {\sc read}($z,y$) message sent by $q_z$.
Thus, it is impossible that $q_z$ receives these messages from strictly
more than $\frac{n}{2}$ processes,
it consequently cannot verify the predicate of line~C\ref{SA-C-05}. It
follows that $q_z$ executes line~C\ref{SA-C-12} with
$my\_view_z[y]=v_y\neq\bot$ and this entails that
$\forall k\in\Pi~:~all\_views_k[z]\neq\bot \Rightarrow
all\_views_k[z][y]\neq\bot$.

Since $z\in min\_\sigma_i$, $all\_views_i[z]\neq\bot$,
$all\_views_i[z][y]\neq\bot$.
According to the predicate of line~C\ref{SA-C-16}, this entails that
$y\in min\_\sigma_i$, and since the previous reasoning holds for any
$y\in min\_\sigma_j$, it shows that $min\_\sigma_j\subseteq min\_\sigma_i$.
It follows that, when $q_i$ executes line~C\ref{SA-C-17}, $\forall y\in
min\_\sigma_j~:~all\_views_i[y]\neq\bot$ and,
consequently, $\forall y\in
min\_\sigma_j~:~all\_views_i[y]=all\_views_j[y]$. It entails that if
$|min\_\sigma_j|<|min\_\sigma_i|$,
then $min\_\sigma_j$ would have been chosen by $q_i$ at
line~C\ref{SA-C-17}, which proves that $min\_\sigma_i=min\_\sigma_j$
and contradicts the fact that $q_i$ and $q_j$ decide differently.
\renewcommand{\toto}{lemma:safe-agr-agreement}
\end{proofL}

\begin{lemma}
\label{lemma:safe-agr-term-decide}
If no simulator crashes while executing  ${\sf propose}()$, then any
invocation of ${\sf decide}()$ by a correct simulator terminates.
\end{lemma}

\begin{proofL}
If no simulator crashes while executing  ${\sf propose}()$,
it follows from Lemma~\ref{lemma:safe-agr-term-propose}
that every simulator $q_i$ that invokes  ${\sf propose}()$ broadcasts
a message  {\sc value} $(i,v_i)$ at line~C\ref{SA-C-01} and
a message  {\sc view} $(i,my\_views_i)$ at line~C\ref{SA-C-12}.

Assuming no  simulator  crashes while executing  ${\sf propose}()$,
let $P$ be the set of simulators that invoke ${\sf propose}()$, and
suppose that one of them, $q_i$, invoke ${\sf decide}()$ and never terminates.
This can only happen if $q_i$ waits forever for the condition of
lines~C\ref{SA-C-15}-C\ref{SA-C-16} to be fulfilled.
Since eventually the messages broadcast by the
simulators of $P$ are all delivered to $q_i$, after some finite time
$\forall y\in P~:~all\_views_i[y]\neq\bot$. Moreover, since
the views broadcast by the  simulators of $P$ are built at
line~C\ref{SA-C-08} from the messages {\sc value}~($-$,$-$)
they receive, it follows that these views can contain non-$\bot$ values
only for the entries corresponding to the simulators of $P$ (the simulators
that are not in $P$ do not sent  messages {\sc value}($-$,$-$)).
Consequently, $p_i$ eventually verifies $\forall y\in
P~:~(all\_views_i[y]\neq\bot)
\land (\{z\in\Pi~:~all\_views_i[y][z]\neq\bot\}\subseteq P)$.
It follows that the property of lines~C\ref{SA-C-15}-C\ref{SA-C-16} is
eventually true for $\sigma=P$, which contradicts the fact that $q_i$
never terminates its ${\sf decide}()$ operation.
\renewcommand{\toto}{lemma:safe-agr-term-decide}
\end{proofL}

\begin{theorem}
\label{theorem-safe-agreement}
The algorithm in Figure~{\em\ref{algo-SA-msg-passing-crash}}
implements a safe agreement object in $\CM[t<n/2]$.
\end{theorem}

\begin{proofT}
The proof follows from
Lemma~\ref{lemma:safe-agr-term-propose} (Propose-Termination),
Lemma~\ref{lemma:safe-agr-validity} (Validity),
Lemma~\ref{lemma:safe-agr-agreement} (Agreement),  and
Lemma~\ref{lemma:safe-agr-term-decide} (Decide-Termination).
\renewcommand{\toto}{theorem-safe-agreement}
\end{proofT}

\subsection{Simulation algorithm}

The simulation algorithm takes as input a distributed algorithm $A$
solving a (colorless) task in the system model $\CMprim[t<n']$, and
simulates it in  $\CM[t<n/2]$.  Each simulator $q_i$, $1\leq i\leq n$,
is given a copy  of the $n'$ processes of $A$, and a private input vector
$input_i[1..n']$, with one input per simulated processes $p_j$.

The simulation consists in a fair simulation by each of the $n$
simulators $q_i$ of the $n'$ simulated processes $p_j$.
To that end,  each simulator manages $n'$ threads
(each simulating a process $p_j$), and the $n$ threads
associated with the simulation of a process $p_j$ cooperate
through safe agreement objects.

\paragraph{Objects shared by the simulators}
To produce a consistent simulation, for each simulated process $p_j$,
the $n$ simulators have to agree on the  same sequence of the messages
received by $p_j$. To that end, they use an array of safe agreement objects,
denoted $\SA[1..n',-]$, such that  $\SA[j,sn]$ allows them to agree on the
$sn$-th message received by the $n'$  threads simulating $p_j$ at each
simulator $q_i$.

\paragraph{Objects managed by each simulator $q_i$}
Each simulator manages the following data structures, with respect to each
simulated process $p_j$.
\begin{itemize}
\vspace{-0.2cm}
\item
$input_i[j]$ contains the input of the simulated process $p_j$,
proposed by the simulator $q_i$. (Simulators are allowed to propose
different input vectors for the simulated processes).
\vspace{-0.2cm}
\item
$sn_i[j]$ is the sequence number (from the simulation point of view) of the
next message received by the simulated process $p_j$.
\vspace{-0.2cm}
\item
$sent_i[j]$ is  a sequence containing  messages sent by the simulated
processes  to the simulated process $p_j$. It is assumed that the $n'$
threads of $q_i$ access $sent_i[j]$ in mutual exclusion
(when they add messages to or  withdraw  messages from this sequence).
The symbol $\oplus$ is used to add messages at the end of a sequence.
Sometimes $sent_i[j]$ is used as a set.
\vspace{-0.2cm}
\item
$received_i[j]$ is  a set containing the messages received by the simulated
process $p_j$ (init.~$\emptyset$).
\vspace{-0.2cm}
\item
$state_i[j]$ contains the current local state of the simulated process
$p_j$. $input_i[j]$ is a part of $state_i[j]$.

It is assumed that the behavior of each simulated process $p_j$ 
is described by a deterministic
transition function $\delta_j()$, such that  $\delta_j(state_i[j],msg)$
(a) simulates $p_j$ until its next message reception, and (b)
returns a pair. This pair is made up of the new local state of $p_j$ plus
an array $msgs[1..n']$ where  $msgs[x]$ contains  messages sent by $p_j$
to the simulated process $p_x$.
\end{itemize}

In addition to the previous local data, each simulator  $q_i$ uses a
starvation-free
mutual exclusion lock, whose operations are denoted ${\sf mutex\_in}_i()$
and ${\sf mutex\_out}_i()$. This lock is used to ensure that, at any time,
at most one of the $n'$ threads of $q_i$ access a safe agreement object.
This is to guarantee that the crash of a simulator $q_i$  entails the crash
of {\it at most one} simulated process $p_j$ (line~\ref{Simu-C-09}).
More precisely, if $q_i$ crashes while executing $\SA[j,sn].{\sf propose}.()$,
it can block forever  only the invocations of  $\SA[j,sn].{\sf decide}.()$,
issued by the other simulators, thereby preventing  the simulation of $p_j$
from terminating.

\renewcommand{\linenumbering}{\ifthenelse{\value{linecounter}<10}
{(0\arabic{linecounter})}{(\arabic{linecounter})}}
\begin{figure}[th!]
\begin{myalgo}
\resetline
\line{Simu-C-01}
\> ${\sf mutex\_in}_i();\SA[j,0].{\sf propose}.(input_i[j]);
    {\sf mutex\_out}_i();$\\

\line{Simu-C-02}
\> $input_i[j] \leftarrow \SA[j,0].{\sf decide}()$;\\

\line{Simu-C-03}
 \>  $\langle state_i[j], msgs[1..n']\rangle \leftarrow
               \delta_j(state_i[j], \emptyset)$;\\

\line{Simu-C-04}
 \>  {\bf for each} $x\in \{1,...,n'\}$  {\bf do}
 $sent_i[x] \leftarrow sent_i[x] ~\oplus~ msgs[x]$  {\bf end for}; \\

\line{Simu-C-05}
\> $sn_i[j] \leftarrow 0$;\\

\line{Simu-C-06}
\> {\bf repeat forever}\\

\line{Simu-C-07}
\> \> $sn_i[j] \leftarrow sn_i[j] +1$;\\

\line{Simu-C-08}
\> \> {\bf wait} $\big((sent_i[j]\setminus received_i[j])\neq \emptyset\big)$;\\

\line{Simu-C-09}
\> \> $msg  \leftarrow \mbox{ oldest message in }
             sent_i[j]\setminus received_i[j] $;\\

\line{Simu-C-10}
\> \> ${\sf mutex\_in}_i()$; $\SA[j,sn_i[j]].{\sf propose}.(msg);$
      ${\sf mutex\_out}_i()$;\\

\line{Simu-C-11}
\> \> $rec\_msg \leftarrow \SA[j,sn_i[j]].{\sf decide}();$\\

\line{Simu-C-12}
\> \> $received_i[j] \leftarrow received_i[j] \cup \{rec\_msg\}$;\\

\line{Simu-C-13}
\> \>      $\langle state_i[j], msgs[1..n']\rangle \leftarrow
               \delta_j(state_i[j], rec\_msg)$;\\

\line{Simu-C-14}
\> \>  {\bf for each} $x\in \{1,...,n'\}$  {\bf do}
 $sent_i[x] \leftarrow sent_i[x] ~\oplus~ msgs[x]$  {\bf end for}; \\

\line{Simu-C-15}
\> \>  {\bf if} (no value yet decided by $p_j$ $\wedge$~
                    $ state_i[j]$ allows $p_j$ to decide a value $v$)  \\

\line{Simu-C-16}
\> \> \>  {\bf then}  the simulated process $p_j$ decides  $v$\\

\line{Simu-C-17}
\> \>  {\bf end if} \\

\line{Simu-C-18}
\> {\bf end repeat}.

\end{myalgo}
\caption{Thread of the simulator  $q_i$, $1\leq i\leq n$, simulating
the process $p_j$, $1\leq j\leq n'$}
\label{algo:simulation-crash}
\end{figure}

\paragraph{The simulation algorithm}
The algorithm describing the simulation of a process $p_j$
by the associated thread of the simulator $q_i$ is presented in
Figure~\ref{algo:simulation-crash}.

The simulators have first to agree on the same input for process $p_j$.
To this end, they use the safe agreement object $\SA[j,0]$
(lines~\ref{Simu-C-01}-\ref{Simu-C-02}).  Moreover, when considering all
the simulated processes, it follows from the
mutual exclusion lock that, whatever the number of simulated processes,
a simulator $q_i$ is engaged in at most one
invocation of ${\sf propose}()$ at a time.
Then, according to the decided  input of $p_j$, $q_i$ locally simulate $p_j$
until it invokes a message  reception (lines~\ref{Simu-C-03}-\ref{Simu-C-04}).

After this initialization, each simulator $q_i$ enters a loop whose aim
is to locally  simulate $p_j$. To this end, $q_i$ first determines
the message that $p_j$ will receive; this message is saved in $rec\_msg$
and added to $received_i[j]$ (lines~\ref{Simu-C-07}-\ref{Simu-C-12}).
When this message has been determined, $q_i$ simulates the behavior of $p_j$
until its next message reception  (lines~\ref{Simu-C-13}-\ref{Simu-C-14}).
Finally, if $state_i[j]$ allows $p_j$ to decide a value with respect to
the simulated decision task, this value is decided
(lines~\ref{Simu-C-15}-\ref{Simu-C-17}).

\subsection{Proof of the simulation}
The reader interested in a formal definition of the term {\it simulation}
--as used here-- will consult~\cite{BGLR01}.

\begin{lemma}
\label{lemma:at-most-one-crash}
The crash of a simulator $q_i$ entails the crash of at most one simulated
 process $p_j$.
\end{lemma}

\begin{proofL}
The only places where a simulator $q_i$ can block is during the invocation of
the safe agreement operation ${\sf decide}()$. Such invocations appear at
line~\ref{Simu-C-02}, and line~\ref{Simu-C-11}. It follows from the
termination property of the safe agreement objects that such an invocation
can block forever the invoking process only if a simulator crashes
during the invocation of  the operation ${\sf propose}()$ on the same object.
But, due to the mutual exclusion lock used at line~\ref{Simu-C-01} and
line~\ref{Simu-C-10}, a simulator can be engaged in at most one
invocation of propose at a time. It follows that the crash of a simulation
$q_i$ can entail the definitive halting (crash) of at most one simulated
process $p_j$.
\renewcommand{\toto}{lemma:at-most-one-crash}
\end{proofL}

\begin{lemma}
\label{lemma:same-seq-of-messages}
The simulation of the reception of the $k$-th
message received by a simulated process $p_j$, returns the same message
at all simulators.
\end{lemma}

\begin{proofL}
The simulation of the message receptions for a simulated process $p_j$,
are executed at each simulator $q_i$ at lines~\ref{Simu-C-08}-\ref{Simu-C-11},
and all the simulators use the same sequence of sequence numbers
(line~\ref{Simu-C-07}).
It then follows from the agreement property of the safe agreement object
$\SA[j,sn]$, that no two simulators obtain different messages
when they invoke  $\SA[j,sn].{\sf decide}()$, and the lemma follows.
\renewcommand{\toto}{lemma:same-seq-of-messages}
\end{proofL}

\begin{lemma}
\label{lemma:one-dec-value}
For every simulated processes $p_j$, no two simulators return
different values.
\end{lemma}

\begin{proofL}
The only non-deterministic elements of the simulation are the input vectors
$input_i[1..n']$ at each simulator $q_i$, and the reception of the simulated
messages.

The lines~\ref{Simu-C-01}-\ref{Simu-C-02} of the
simulation force  the simulators to agree on the same input value
for each simulated process $p_j$, $1\leq j \leq n'$.
Similarly,  as shown by Lemma~\ref{lemma:same-seq-of-messages},
for each simulated process $p_j$, the lines~\ref{Simu-C-07}-\ref{Simu-C-11}
direct the simulators to  agree on the very same sequence of messages
received by $p_j$. It follows from the fact that the function $\delta_j()$
is deterministic, that   any two simulators $q_i$ and $q_k$,
that execute  lines~\ref{Simu-C-15}-\ref{Simu-C-16} during the
same ``round number'' $sn_i[j]=sn_k[j]$, are such that $state_i[j]=state_k[j]$,
from which the lemma follows.
\renewcommand{\toto}{lemma:one-dec-value}
\end{proofL}

\begin{lemma}
\label{lemma:correct-simulation}
The sequences of message receptions simulated by each simulator $q_i$
on behalf of each simulated process $p_j$, define a correct execution of
the simulated algorithm.
\end{lemma}

\begin{proofL}
To prove the correctness of the simulation, we have  to show that
\begin{enumerate}
\vspace{-0.2cm}
\item
\label{item1}
Every message that was sent by a simulated process to another simulated
process (whose simulation is not blocked either), is received, and
\vspace{-0.2cm}
\item
\label{item2}
The simulated messages respect a simulated physical order
(i.e., no message is ``received'' before being  ``sent'').
\end{enumerate}

Item~\ref{item1} is satisfied because the messages sent by
the simulated process $p_j$ to the simulated process $p_k$ are received
(lines~\ref{Simu-C-09}-\ref{Simu-C-11}) in their sending order
(as defined at line~\ref{Simu-C-04}  and line~\ref{Simu-C-14}).
Hence, if $p_k$ is not blocked (due to the crash of a simulator)
it obtains the messages from $p_j$ in their sending order.

For Item~\ref{item2}, let us define a (simulated) physical order as follows.
For each simulated message $m$, let us consider the first time at which the
reception of $m$  was simulated (i.e., this occurs when
 --for the first time-- a simulator terminates  the invocation  of
$\SA[-,-].{\sf decide}()$  that returns $m$).
A message that is decided has been proposed by a simulator to a safe agreement
object before  being decided (validity property).The  sending time of  a
simulated message is defined as  the first time at which
$\SA[-,-].{\sf propose}(m)$ is  invoked by a simulator.
It follows that any simulated  message is sent before being received,
which concludes the lemma.
\renewcommand{\toto}{lemma:correct-simulation}
\end{proofL}

\begin{lemma}
\label{lemma:nb-simulated-proc}
Each correct simulator $q_i$ computes the decision value of
at least $(n'-t)$ simulated processes.
\end{lemma}

\begin{proofL}
Due to Lemma~\ref{lemma:at-most-one-crash}, and the fact that at most $t$
simulators may crash, it follows that at most $t$ simulated processes may be
prevented from progressing. As (a) by assumption the simulated algorithm $A'$
is $t$-resilient, and (b) due to Lemma~\ref{lemma:correct-simulation}
the simulation produces a correct simulation of $A'$, it follows that at
least $(n'-t)$ simulated processes decide a value.
\renewcommand{\toto}{lemma:nb-simulated-proc}
\end{proofL}

\begin{theorem}
\label{theorem:main-simulation}
Let $A$ be an algorithm solving a decision task in $\CMprim[t<n']$.
The algorithm described in Figure~\ref{algo:simulation-crash} is a
correct simulation of $A$ in~$\CM[t<n/2]$.
\end{theorem}

\begin{proofT}
The theorem follows from  Lemma~\ref{lemma:correct-simulation} and Lemma~\ref{lemma:nb-simulated-proc}.
\renewcommand{\toto}{theorem:main-simulation}
\end{proofT}

\section{BG(MP,B): BG in the Byzantine Asynchronous Message-Passing  Model}
\label{sec:BG-byzantine-model}

This section presents an algorithm, denoted BG(MP,B), which implements the BG
simulation in the Byzantine asynchronous message-passing model $\BM[t<n/3]$.
To this end, an appropriate  safe agreement object is first built, and then
used by the simulation algorithm.

\subsection{From crash failures to Byzantine behaviors}
The idea is to extend the algorithm of Figure~\ref{algo-SA-msg-passing-crash}
to obtain an algorithm that copes with Byzantine simulators.
The main issues that have to be solved are the following.
\begin{itemize}
\vspace{-0.2cm}
\item
The simulators  need a  mechanism to  control the  validity of the inputs
to the safe agreement objects. (See below for the notion of a valid value.)
\vspace{-0.2cm}
\item
The simulators must be able to check if a given simulator $q_i$ is
participating in more than one operation ${\sf propose}()$ at the same time
(on the same or several safe agreement objects).
If it is the case, $q_i$ is faulty and its definitive stop  can block forever
several simulated processes. Hence, such a faulty simulator has to be ignored.
\end{itemize}
To solve these issues, each safe agreement object may no longer be considered
as a separate abstraction: each new instance depends on the previous ones. This
is captured in the following specification customized to the Byzantine
model, and, at the operational level, in the predicate ${\sf valid}()$
used in the algorithm implementing the operation ${\sf propose}()$.

\subsection{Safe agreement in  $\BM[t<n/3]$: definition}
\label{sec:spec-Byzantine-SA}
To cope with  the previous observations, the fact that
a faulty process may decide an arbitrary  value,
and the fact  that the safe agreement objects are used  to solve
specific problems (a simulation in our case),
the specification of the safe agreement object is reshaped as follows.

A value proposed by a process to a safe agreement object must be
{\it valid}. At each correct simulator $q_i$, the validity of a value is 
captured by a predicate denoted ${\sf valid}_i(j,v)$
where $v$ is the value and $q_j$ the simulator that proposed it.
This predicate is made up of two parts (defined in Section~\ref{sec:is-valid} 
and  Section~\ref{sec:simulation-byzantine}, respectively).
If $q_j$ is correct, the predicate ${\sf valid}_i(j,v)$ eventually 
returns $\mathit{true}$ at $p_i$. If
$q_j$ is faulty,  ${\sf valid}_i(j,v)$  returns  $\mathit{true}$ at $p_i$
only if
 (a) the value $v$  could have been proposed by a correct simulator and (b) 
 to $q_i$'s knowledge, $q_j$ does not participate concurrently in several 
invocations of  ${\sf propose}()$.
\vspace{-0.1cm}
\begin{itemize}
\item Validity.
If a correct simulator $q_i$ decides the value $v$, there is 
a correct simulator $q_j$ such that ${\sf valid}_j(-,v)$.
($v$ was validated by a correct simulator.)
\vspace{-0.2cm}
\item Agreement. No two correct simulators decide distinct values.
\vspace{-0.2cm}
\item Propose-Termination.
Any invocation of  ${\sf propose}()$ by a correct simulator terminates.
\vspace{-0.2cm}
\item Decide-Termination.
The invocations by all the correct simulators  of ${\sf  decide}()$
on all the safe agreement objects terminate, except for at most $t$
safe agreement objects.
\end{itemize}

\subsection{Safe agreement in  $\BM[t<n/3]$: algorithm}
\label{sec:is-valid}

The local variables $values_i[1..n]$, $my\_view_i[1..n]$,
$all\_views_i[1..n]$,  and the algorithm implementing the operation
${\sf decide}()$ are the same as in Figure~\ref{algo-SA-msg-passing-crash}
(lines~C\ref{SA-C-15}-C\ref{SA-C-19}).
The  new algorithm implementing the  operation ${\sf  propose}()$, and the
processing of the associated messages, are  described in
Figure~\ref{algo-SA-msg-passing-byzantine-propose} and
Figure~\ref{algo-SA-msg-passing-byzantine-answers}.

This implementation uses an additional local array
$answers_i[1..n][1..n][1..n]$, all entries of which are initialized to
``?''. The meaning of  ``$answers_i[k][j][x]=v$'' 
(where $v$ is a proposed value or $\bot$) is the following:
to the knowledge of $q_i$, the simulator $q_k$ answered  value $v$ when
it received  the message {\sc read}$(j,x)$ sent by $q_j$.
(A simulator $q_j$ broadcasts such a message when it needs to know the value
proposed by  the simulator $q_x$; $\bot$ means that $q_k$ does not 
know this value yet.) This  means that, from  $q_i$'s point of
view, the value proposed by $q_x$, as known by $q_k$ when it received 
the request by $q_j$, is $v$.

\begin{lemma}
\label{lemma:quorum-intersection}
Any two sets of simulators $Q_1$ and $Q_2$ of more than $\frac{n+t}{2}$
elements have at least one correct simulator in their intersection.
\end{lemma}

\begin{proofL}
As  we   consider  integers,  ``strictly  more  than  $\frac{n+t}{2}$''
is equivalent to ``at least $\lfloor\frac{n+t}{2}\rfloor+1$''.

 \begin{itemize}
\vspace{-0.2cm}
\item
$Q_1\cup Q_2\subseteq \{p_1,\ldots, p_n\}$. Hence, $|Q_1 \cup Q_2|\leq n$.
\vspace{-0.2cm}
\item
$|Q_1\cap Q_2|=|Q_1|+|Q_2|-|Q_1\cup Q_2|\ge |Q_1|+|Q_2|-n \ge
     2(\lfloor\frac{n+t}{2}\rfloor+1)-n>2(\frac{n+t}{2})-n=t$.
Hence,  $|Q_1\cap Q_2|\ge t+1$. It follows that  $Q_1\cap Q_2$
contains at least one correct simulator.
\end{itemize}
\vspace{-0.4cm}
\renewcommand{\toto}{lemma:quorum-intersection}
\end{proofL}

The fact that, despite Byzantine processes,  the intersection of any two
simulator  sets of  size greater than $\frac{n+t}{2}$ have at least one
correct simulator in common, is used in many places in the algorithm.
This property will be used in the proof to show that the local views
of the correct processes are mutually consistent.

\paragraph{The operation ${\sf propose}()$}
The client side of the algorithm implementing the operation ${\sf propose}()$
is described in Figure~\ref{algo-SA-msg-passing-byzantine-propose}; its server
side is described in Figure~\ref{algo-SA-msg-passing-byzantine-answers}.
The client side  algorithm is very close to the one of the crash failure case
(Figure~\ref{algo-SA-msg-passing-crash}).  They differ in two points.
\begin{itemize}
\vspace{-0.2cm}
\item
The  message tags {\sc value} and {\sc view} (used at
lines~C\ref{SA-B-02},~C\ref{SA-B-06},  and~C\ref{SA-B-13}
in Figure~\ref{algo-SA-msg-passing-crash}) are replaced
in Figure~\ref{algo-SA-msg-passing-byzantine-propose}
by the  tags  {\sc value'ack} and {\sc view'ack}, respectively. 
The role of these message tags is explained below.
\vspace{-0.2cm}
\item
The predicate of line~B\ref{SA-B-05}  is replaced by the predicate
$|\{k~:~answers_i[k][i][x]=\bot\}|>\frac{n+t}{2}$.
This predicate states that more than $\frac{n+t}{2}$ simulators
answered $\bot$  to the request message {\sc  read}$(i,x)$ broadcast by $q_i$,
(i.e.,  they  did   not know the  value   proposed   by $q_x$   when  they
received the  read request).
\end{itemize}

\renewcommand{\linenumbering}{\ifthenelse{\value{linecounter}<10}
{(B0\arabic{linecounter})}{(B\arabic{linecounter})}}
\begin{figure}[ht!]
\begin{myalgo}
\resetline
{\bf operation} $\mathsf{propose}$ ($v_i$) {\bf is}\\
\line{SA-B-01}~~ \= $\mathsf{broadcast}$ {\sc value} $(i,v_i)$;\\
\line{SA-B-02}    \>   {\bf    wait}    \big({\sc   value'ack}    $(i,v_i)$
$\mathsf{received}$  from  $> \frac{n+t}{2}$  different
simulators\big);\\

\line{SA-B-03}    \>   {\bf    for   each}    $x\in   [1..n]$    {\bf   do}
$\mathsf{broadcast}$ {\sc read} $(i,x)$ {\bf end for};\\

\line{SA-B-04} \> {\bf for each} $x\in [1..n]$ {\bf do}\\

\line{SA-B-05}          \>      \>            \>
 {\bf wait}
  \big(\=($|\{k~:~answers_i[k][i][x]=\bot\}|>\frac{n+t}{2}$) $\lor$\\

\line{SA-B-06}   \> \> \>  \>($~\exists~w:$   {\sc    value'ack}   $(x,w)$
$\mathsf{received}$    from  $> \frac{n+t}{2}$ different simulators)\big);\\

\line{SA-B-07}  \>   \> \> 
{\bf if} (predicate of line~B\ref{SA-B-06} satisfied)\\
\line{SA-B-08}  \>   \> \> \> {\bf then} \= $my\_view_i[x] \gets w$\\

\line{SA-B-09}   \>  \> \> \> {\bf else} \> $my\_view_i[x] \gets \bot$\\

\line{SA-B-10}  \>   \> \> {\bf end if}\\
\line{SA-B-11} \> {\bf end for};\\

\line{SA-B-12} \> $\mathsf{broadcast}$ {\sc view} $(i,my\_view_i)$;\\

\line{SA-B-13}   \>  {\bf   wait}   \big({\sc  view'ack}   $(i,my\_view_i)$
$\mathsf{received}$  from  $> \frac{n+t}{2}$  different simulators\big);\\

\line{SA-B-14} \> ${\sf return}()$.\\~\\

{\bf operation} $\mathsf{decide}$ () {\bf is}\\
(C\ref{SA-C-15})
\>  {\bf wait}
       \big($\exists$ a non-empty set $\sigma \subseteq \Pi$:\\

(C\ref{SA-C-16})
\> \> ~$\forall~y\in \sigma:~
       \big[ (all\_views_i[y]\neq\bot) ~\wedge~
 \big(\forall~z\in \Pi:~(all\_views_i[y][z]\neq\bot)
                               \Rightarrow(z\in \sigma)\big)\big]$;\\

(C\ref{SA-C-17})
\> {\bf let} $min\_\sigma_i$ {\bf be} the set  $\sigma$ of smallest size; \\

(C\ref{SA-C-18})
\>  {\bf let} $res$ {\bf be} $\min(\{values_i[y] ~:~ y\in min\_\sigma_i\})$;\\

(C\ref{SA-C-19})
\> ${\sf return}(res)$.
\end{myalgo}
\caption{Safe agreement object in $\BM[t<n/3]$:
operation  $\mathsf{propose}()$ of simulator $q_i$}
\label{algo-SA-msg-passing-byzantine-propose}
\end{figure}

\begin{figure}[ht!]
\begin{myalgo}

{\bf when the message} {\sc value} $(j,v)$ {\bf is}
 $\mathsf{received ~from}~q_j$ {\bf for the first time}:\\
\line{SA-B-15} \= {\bf wait} \big($\mathsf{valid}_i$ ($j,v$)\big);
$\mathsf{broadcast}$ {\sc value'valid} $(j,v)$.\\~\\

{\bf when the message} {\sc value'valid} $(j,v)$ {\bf is}
 $\mathsf{received}$:\\
\line{SA-B-16} \> {\bf if}
  \=\big(({\sc value'valid} $(j,v)$
    $\mathsf{received}$ from  $> \frac{n+t}{2}$ different simulators)  $\land$  ({\sc
value'witness} $(j,-)$ never broadcast)\big)\\
\line{SA-B-17} \> \> {\bf then} $\mathsf{broadcast}$ {\sc value'witness} $(j,v)$ {\bf end if}.\\~\\

{\bf when the message} {\sc value'witness} $(j,v)$ {\bf is}
 $\mathsf{received}$:\\

\line{SA-B-18}   \>   {\bf    if}   \=\big(({\sc   value'witness}   $(j,v)$
$\mathsf{received}$   from  $t+1$   different  simulators)   $\land$  ({\sc
value'witness} $(j,v)$ never broadcast)\big)\\

\line{SA-B-19} \> \>
    {\bf then} $\mathsf{broadcast}$ {\sc value'witness} $(j,v)$\\
\line{SA-B-20} \> {\bf end if};\\
\line{SA-B-21} \> {\bf if}
  \=({\sc value'witness} $(j,v)$
    $\mathsf{received}$ from  $> \frac{n+t}{2}$ different simulators)\\
\line{SA-B-22} \> \> {\bf then}
   \=$values_i[j]\gets v$; $\mathsf{broadcast}$ {\sc value'ack} $(j,v)$\\

\line{SA-B-23} \> {\bf end if}.\\~\\

--------------------------------------------------------------------------------------------------------------------------------------------------\\

{\bf  when  the message}  {\sc  read}  $(j,x)$  {\bf is}  $\mathsf{received
~from}~q_j$ {\bf for the first time}:\\

\line{SA-B-24}    \>    {\bf    wait}    \big({\sc    value'ack}    $(j,v)$
$\mathsf{received}$  from $> \frac{n+t}{2}$  different simulators\big);\\

\line{SA-B-25}      \>       $values_i[j]\gets v$;
$\mathsf{broadcast}$ {\sc value'ack} $(j,v)$;\\

\line{SA-B-26}      \>      $\mathsf{broadcast}$     {\sc      read'answer}
$(j,x,values_i[x])$.\\~\\

{\bf   when   the   message}   {\sc   read'answer}   $(j,x,v)$   {\bf   is}
$\mathsf{received ~from}~q_k$ {\bf for the first time}:\\

\line{SA-B-27}  \> {\bf  if} ({\sc  read'answer'witness}  $(k,j,x,-)$ never
broadcast)   {\bf  then}  $\mathsf{broadcast}$   {\sc  read'answer'witness}
$(k,j,x,v)$ {\bf end if}.\\~\\

{\bf when the message}
{\sc read'answer'witness} $(k,j,x,v)$ {\bf is} $\mathsf{received}$:\\

\line{SA-B-28} \> {\bf if} \=\big(\=({\sc read'answer'witness}
$(k,j,x,v)$ $\mathsf{received}$ from $t+1$ different simulators)\\

\line{SA-B-29} \> \>
$~\land$ ({\sc read'answer'witness} $(k,j,x,v)$ never broadcast)\big)\\
\line{SA-B-30} \> \>
{\bf then} $\mathsf{broadcast}$ {\sc read'answer'witness} $(k,j,x,v)$\\
\line{SA-B-31} \> {\bf end if};\\
\line{SA-B-32} \> {\bf if}
\=({\sc read'answer'witness} $(k,j,x,v)$ $\mathsf{received}$
          from $> \frac{n+t}{2}$ different simulators)\\
\line{SA-B-33} \> \> {\bf then} $answers_i[k][j][x]\gets v$\\
\line{SA-B-34} \> {\bf end if}.\\~\\
--------------------------------------------------------------------------------------------------------------------------------------------------\\
{\bf when the message} {\sc view} $(j,view)$ {\bf is} $\mathsf{received~from}~q_j$ {\bf for the first time}:\\
\line{SA-B-35} \> {\bf if} \=\big(({\sc view'witness} $(j, -)$ never broadcast) $\land$ ($view[j]\neq\bot$)\big)\\
\line{SA-B-36} \> \>{\bf then} \={\bf for} \=$x\in[1..n]$ {\bf do}\\
\line{SA-B-37} \> \> \> \> {\bf if} \=($view[x]\neq\bot$)\\
\line{SA-B-38} \> \>  \> \> \>
   {\bf then} \={\bf  wait} \big({\sc value'ack}
       $(x,view[x])$ $\mathsf{received}$ from $> \frac{n+t}{2}$
       different simulators\big)\\
\line{SA-B-39} \> \> \> \> \>
     {\bf else} \>
     {\bf wait} \big($|\{k~:~answers_i[k][j][x]=\bot\}|>\frac{n+t}{2}$\big)\\
\line{SA-B-40} \> \> \> \>{\bf end if}\\
\line{SA-B-41} \> \> \> {\bf end for};\\
\line{SA-B-42} \> \> \> $\mathsf{broadcast}$ {\sc view'witness} $(j, view)$\\
\line{SA-B-43} \>{\bf end if}.\\~\\

{\bf when the message}
{\sc view'witness} $(j, view)$ {\bf is} $\mathsf{received}$:\\
\line{SA-B-44} \> {\bf if} \=\big(({\sc view'witness} $(j, view)$
         $\mathsf{received}$ from $t+1$ different simulators)
$~\land$ ({\sc view'witness} $(j, view)$ never broadcast)\big)\\

\line{SA-B-45} \> \>
{\bf then} $\mathsf{broadcast}$ {\sc view'witness} $(j, view)$\\
\line{SA-B-46} \> {\bf end if};\\

\line{SA-B-47} \>
     {\bf if} \=({\sc view'witness} $(j, view)$ $\mathsf{received}$
       from $> \frac{n+t}{2}$ different simulators)\\

\line{SA-B-48} \> \>{\bf then} \=$all\_views_i[j]\gets view$;
          $\mathsf{send}$ {\sc view'ack} $(j,view)$ $\mathsf{to}~q_j$ \\

\line{SA-B-49} \> {\bf end if}.

\end{myalgo}
\caption{Safe agreement object in $\BM[t<n/3]$:
server side of simulator $q_i$}
\label{algo-SA-msg-passing-byzantine-answers}
\end{figure}

\paragraph{Messages {\sc value}$()$,  {\sc value'valid}$()$, 
{\sc value'witness}$()$ and {\sc value'ack}$()$}
When a simulator $q_i$ invokes the operation  ${\sf propose}(v_i)$, it first
broadcasts the message {\sc value} $(i,v_i)$, and waits for
$\frac{n+t}{2}$ acknowledgments (messages {\sc value'ack}$(i,v_i)$,
lines~B\ref{SA-B-01}-B\ref{SA-B-02}).
Then, as in the crash failure case (Figure~\ref{algo-SA-msg-passing-crash}),
it builds its  local view of the values proposed to the safe agreement object
(lines~B\ref{SA-B-03}-B\ref{SA-B-11}).  Finally, it sends  its local  view to
all other simulators (lines~B\ref{SA-B-12}-B\ref{SA-B-13}).

On its server side, when a simulator $q_i$ receives a message
{\sc value} $(j,v)$, it first checks if this message is valid
(line~B\ref{SA-B-15}).
If the message is valid, $q_i$ broadcasts (echoes) the message
{\sc value'valid} $(j,v)$ to inform the other simulators
that it agrees to take into account the pair $(j,v)$ (line~B\ref{SA-B-15}).

When the simulator $p_i$  has received the message {\sc value'valid} $(j,v)$
from more than $\frac{n+t}{2}$ simulators, it broadcasts the message 
{\sc value'witness} $(j,v)$ to inform the other processes that 
at least  $\frac{n+t}{2}-t= \frac{n-t}{2}\geq t+1$ correct simulators,
have validated the pair  $(j,v)$.

When $q_i$ has received the message {\sc value'witness} $(j,v)$
from $(t+1)$ simulators (i.e., from at least one correct simulator) it
broadcasts this message, if not yet done (lines~B\ref{SA-B-18}-B\ref{SA-B-20}).
This is to prevent  invocations of  ${\sf propose}()$
from blocking forever (while waiting {\sc value'ack} $(j,v)$
messages at line~B\ref{SA-B-02}, B\ref{SA-B-06}, B\ref{SA-B-24} or
B\ref{SA-B-38}),   because not enough {\sc value'witness} $(j,v)$ messages
have been broadcast\footnote{A similar mechanism is used in~\cite{B87}
to ensure that the proposed reliable broadcast abstraction guarantees
that a  message is received by all or none of the correct processes.}.
Then, if $q_i$  has received the message {\sc value'witness} $(j,v)$
from more than $\frac{n+t}{2}$ simulators,
it takes $v$ into account  (writes it into $values_i[j]$) and sends an
acknowledgment to $q_j$ (lines~B\ref{SA-B-21}-B\ref{SA-B-23}).
The corresponding message {\sc value'ack} $(j,v)$ broadcast by $q_i$
will  also inform the other simulators  that $q_i$ took  into account the
value  $v$   proposed by  $q_j$. Hence,  this message  will help  $q_j$
progress at line~B\ref{SA-B-02}, and all correct simulators  progress at
line~B\ref{SA-B-06}.

\paragraph{First part of the predicate $\mathsf{valid}_i(j,v)$}
\label{sec:def-is-valid}
As already indicated, the aim of this predicate is  to help a simulator $q_i$
detect if the value $v$ proposed by the simulator  $q_j$ is valid.
It is always satisfied when  $q_j$  is correct, and it can return 
${\mathit true}$  or ${\mathit false}$ when $q_j$ is faulty. 
It is made up of two sub-predicates $P1$ and $P2$.

\begin{itemize}
\vspace{-0.2cm}
\item
The first sub-predicate  $P1$
checks if, for the messages {\sc value} $(j,-)$ (from $q_j$) and 
{\sc value'valid} $(j,-)$ (from more than $t+1$ different simulators) 
that $q_i$ has  received for other safe agreement objects,  $q_i$ has also 
received the associated messages {\sc view'witness}~$(j,-)$ 
from at least $(n-t)$ different simulators.
This allows $q_i$ to check if the simulator $q_j$ is not
simultaneously  participating in other invocations of $\mathsf{propose}()$
on  other safe agreement  objects.
\vspace{-0.2cm}
\item
The aim of the second  sub-predicate $P2$ (defined in
Section~\ref{sec:simulation-byzantine} and used in the simulation) is 
to allow the simulators to check that the simulation is consistent.
As  the present section considers safe agreement objects independently 
from its use in the simulation, we consider, for now,
 that $P2$ is always satisfied.
\end{itemize}
If the full predicate $\mathsf{valid}_i(j,v)$ is never satisfied,  
$q_i$ will, collectively with the other correct simulators,
prevent the faulty simulator $q_j$ from progressing
with respect to the corresponding safe agreement object.

\paragraph{Messages {\sc read}$()$, {\sc read'answer}$()$ and
          {\sc read'answer'witness}$()$}
 After the value $v_i$ it proposes to the safe agreement object has been
taken  into  account by $\frac{n+t}{2 }$ simulators,  $q_i$  builds a
local view  of all the values proposed (array $\mathit{my\_view_i}[1..n]$). To this end, as in  the crash
failure model,  $q_i$  sends to each simulator $q_x$ the customized message
{\sc read}~$(i,x)$ (line~B\ref{SA-B-03}).
Its behavior is then similar to the one of the crash failure model
(line~B\ref{SA-B-04}-B\ref{SA-B-11}), where the new predicate
$|\{k~:~answers_i[k][i][x]=\bot\}|>\frac{n+t}{2}$ is now used at
line~B\ref{SA-B-05}.

When  $q_i$ receives  the message   {\sc read}~$(j,x)$  from  the simulator
$q_j$, it first waits until it knows that the  value proposed by $q_j$
is known by more than $\frac{n+t}{2}$ simulators (line~B\ref{SA-B-24}).
This is to check that $q_j$ broadcast its proposed value before
reading the other simulator values used  to build its own view.
When  this   occurs,  $q_i$ answers the  message {\sc  read}~$(j,x)$
by broadcasting the message  {\sc read'answser}~$(j,x,values_i[x])$
 to inform all the simulators on what it currently knows on the value
proposed by $q_x$ (line~B\ref{SA-B-26}). (Let us remind that, in  the
crash failure model, $q_i$ was sending this message only to $q_j$.)

When it receives the message {\sc read'answer}~$(j,x,v)$ from a simulator
$q_k$,  if not yet done,   $q_i$ broadcasts the message
{\sc read'answser'witness}~$(k,j,x,v)$.
The lines~B\ref{SA-B-27}-B\ref{SA-B-31} implement a reliable broadcast
\cite{B87}, i.e., the message {\sc read'answser'witness}~$(k,j,x,v)$
is   received by all correct  processes  or none of  them,  and is  always
received if the sender is correct. The reliable  reception of  this
message entails the assignment of $answer_i[k,j,x]$ to $v$
(line~B\ref{SA-B-33}).

\paragraph{Messages {\sc view}$()$, {\sc view'witness}$()$
           and {\sc view'ack}$()$}
Finally, as in Figure~\ref{algo-SA-msg-passing-crash},
the simulator $q_i$ broadcasts its local view of proposed values to
all  simulators, waits until more than  $\frac{n+t}{2}$ of them sent back an
acknowledgment, and returns  from  the  invocation  of
${\sf propose}()$ (lines~B\ref{SA-B-12}-B\ref{SA-B-14}).

When $q_i$ receives for the first time the message  {\sc view}~$(j,view)$,
it realizes an enriched reliable broadcast whose aim is to assign
$view$ to $all\_view_i[j]$. Let us first observe that if $view[j]=\bot$,
then $q_j$ is Byzantine. If it has not yet broadcast
{\sc  view'witness}~$(j,view)$ and  if $view[j]\neq\bot$
(line~B\ref{SA-B-35}), $q_i$ first
checks if all the  values in  $view[1..n]$ are  consistent.
From its  point of view, this means that, for each simulator $q_x$,
(a) if $view[x]=v$, it must  receive messages {\sc value'ack}~$(x,v)$
from more than $\frac{n+t}{2}$ simulators, and (b) if  $view[x]=\bot$,
the same predicate as in line~B\ref{SA-B-05} must become satisfied.
This consistency check is realized by the lines~B\ref{SA-B-36}-B\ref{SA-B-41}.

Finally, when  $q_i$ receives a message  {\sc view'witness}~$(j,view)$,
it does the following. First, if  it has  received this message from at
least one correct simulator, and has not yet broadcast it, $q_i$ does it
(lines~B\ref{SA-B-44}-B\ref{SA-B-46}). This part of the reliable broadcast is
to prevent the correct simulators from blocking forever.
Then, if it has received  {\sc view'witness}~$(j,view)$ from more than
 $\frac{n+t}{2}$ simulators and  has not yet assigned a value to
$all\_view_i[j]$, $q_i$ does it and sends to $q_j$ the acknowledgment
message {\sc view'ack}~$(j,view)$ to inform $q_j$  that it knows its view
(lines~B\ref{SA-B-47}-B\ref{SA-B-49}).

\subsection{A communication pattern}
When considering the algorithm of
Figure~\ref{algo-SA-msg-passing-byzantine-answers}, it appears
that the  processing of the messages 
{\sc value'witness}~$()$  (lines~B\ref{SA-B-18}-B\ref{SA-B-23}), 
{\sc read'answer'witness}~$()$ (lines~B\ref{SA-B-28}-B\ref{SA-B-34}), and
{\sc view'witness}~$()$ (lines~B\ref{SA-B-44}-B\ref{SA-B-49}), 
follow the same generic pattern. 
This pattern, inspired from~\cite{B87} and 
where {\sc witness} is used as message tag, 
is described in Figure~\ref{algo:generic-pattern}.

\renewcommand{\linenumbering}{\ifthenelse{\value{linecounter}<10}
{(GP0\arabic{linecounter})}{(GP\arabic{linecounter})}}
\begin{figure}[ht]

\centering{\fbox{
\begin{minipage}[t]{150mm}
\footnotesize
\renewcommand{\baselinestretch}{2.5}
\resetline
\begin{tabbing}
aaaaaaaaa\=aaa\=aaaaa\=aaaaaa\=\kill
{\bf when} {\sc witness} $(m)$ {\bf is} $\mathsf{received}$:\\

\line{GP-01}  \> 
{\bf if} \=\big(\={\sc witness} $(m)$ $\mathsf{received}$ 
                                   from $t+1$ different simulators\\

\line{GP-02} \> \>
$~\land$ {\sc witness} $(m)$ never broadcast\big)\\

\line{GP-03} \> \>
{\bf then} $\mathsf{broadcast}$ {\sc witness} $(m)$\\

\line{GP-04} \> {\bf end if};\\

\line{GP-05} \> {\bf if}
\=({\sc witness} $(m)$ $\mathsf{received}$
          from $> \frac{n+t}{2}$ different simulators)\\

\line{GP-06} \> \> {\bf then} execute statement $A$\\

\line{GP-07} \> {\bf end if}.

\end{tabbing}
\end{minipage}
}
\caption{Generic communication pattern in $\BM[t<n/3]$}
\label{algo:generic-pattern}
}
\end{figure}

\begin{theorem}
\label{theo-generic-pattern}
{\em (i)} If a correct simulator  executes  action $A$,
all correct simulators do it.\\
{\em(ii)}  If $(t+1)$ correct simulators execute ${\sf broadcast}$
{\sc witness}$(m)$,  all correct simulators execute action $A$.
\end{theorem}

\begin{proofT}
Proof of (i). Let $p_i$ be a correct process that executes $A$.
It follows from line GP\ref{GP-05} that it has received the message
{\sc witness} $(m)$ from more than $\frac{n+t}{2}$ different simulators.
As $n>3t$, $\lfloor \frac{n+t}{2}\rfloor +1 \geq 2t+1$,
$p_i$ received the message {\sc witness} $(m)$ from at least $(t+1)$
correct simulators. It then follows from lines~GP\ref{GP-01}-GP\ref{GP-02}
that all correct simulators broadcast {\sc witness} $(m)$ and, consequently,
each correct simulator receives {\sc witness} $(m)$ from at least $(n-t)$
simulators. The proof follows from $n-t > \frac{n+t}{2}$.

Proof of (ii). If $(t+1)$ correct simulators broadcast {\sc witness} $(m)$,
the predicate of line~GP\ref{GP-01} is eventually  satisfied
at every correct simulator.  As  As $n-t> \frac{n+t}{2}$,
it follows that the predicate of line~GP\ref{GP-05}  will also be
satisfied at each correct simulator, which concludes the proof.
\renewcommand{\toto}{theo-generic-pattern}
\end{proofT}
\subsection{Safe agreement object in $\BM[t<n/3]$: proof}
This section proves that the algorithm presented in
Figures~\ref{algo-SA-msg-passing-byzantine-propose}
and~\ref{algo-SA-msg-passing-byzantine-answers} implements a
safe agreement object in the presence of Byzantine simulators,
i.e., any of its runs in  $\BM[t<n/3]$  satisfies the validity, agreement,
and termination properties that define this object.

\paragraph{Propose-termination}

\begin{lemma}
\label{lemma-termination-if-valid}
Let $q_i$ be a correct simulator.
If the predicate ${\sf valid}_j(i,v_i)$ eventually becomes satisfied
at the correct simulators $q_j$,
then the invocation of ${\sf propose}(v_i)$ by $q_i$ terminates.
\end{lemma}

\begin{proofL}
A correct simulator $q_i$ can be blocked forever in a ${\sf wait}$ statement
(1) at line B\ref{SA-B-02},  (2) at lines B\ref{SA-B-05}-B\ref{SA-B-06}, or
(3) at line B\ref{SA-B-13}. We show that, if the predicate
${\sf valid}_j(i,v_i)$  is eventually  satisfied at the correct simulators
$q_j$,  $p_i$ cannot block forever in the invocation of ${\sf propose}(v_i)$.

\begin{itemize}
\vspace{-0.2cm}
\item ${\sf wait}$ instruction at line B\ref{SA-B-02}. \\
Simulator $q_i$ first broadcasts the message {\sc value}$(i,v_i)$
(line B\ref{SA-B-01}), then waits for {\sc value'ack} messages from more than
$\frac{n+t}{2}$ different simulators.
When a correct simulator $q_j$ receives {\sc value}$(i,v_i)$ for the
first time, it waits until ${\sf valid}_j(i,v_i)$  becomes satisfied.
By assumption, this  happens.  Simulator $q_j$  then broadcasts
{\sc value'valid}$(i,v_i)$. It follows that each of the at least $(n-t)$
correct simulators  broadcasts the message {\sc value'valid}$(i,v_i)$.

As  $n-t>\frac{n+t}{2}$, it follows that each
correct simulator  $q_j$ receives the message {\sc value'valid}$(i,v_i)$
from more than $\frac{n+t}{2}$ simulators and broadcasts the message
{\sc value'witness}$(i,v_i)$.

 According to Theorem~\ref{theo-generic-pattern}, $q_j$ updates $values_j[i]$
 with  $v_i$, and broadcasts {\sc value'ack}$(i,v_i)$
(lines~B\ref{SA-B-21}-B\ref{SA-B-23}).
The correct simulator $q_i$ will then receive the message
{\sc value'ack}$(i,v_i)$ from at least  $n-t > \frac{n+t}{2}$
simulators. Hence, it cannot block forever at line  B\ref{SA-B-02}.

\vspace{-0.2cm}
\item ${\sf wait}$ instruction at lines B\ref{SA-B-05}-B\ref{SA-B-06}.\\
In this waiting statement, $q_i$ waits until
either $|\{k~:~answers_i[k][i][j]=\bot\}|>\frac{n+t}{2}$ becomes true, or
until it receives {\sc value'ack}$(j,w)$ from more than $\frac{n+t}{2}$
different simulators.
\vspace{-0.3cm}
\begin{itemize}
\item
If $q_j$ is a correct simulator that  invoked ${\sf propose}(j,w)$,
the reasoning is the same as above. Consequently, $q_i$ will receive
{\sc value'ack}$(j,w)$ from at least $n-t>\frac{n+t}{2}$ different
simulators.
\vspace{-0.1cm}
\item
If $q_j$ is faulty or never invokes ${\sf propose}(j,w)$, $q_i$ may never
receive {\sc value'ack}$(j,w)$ from more than $\frac{n+t}{2}$ different
simulators. We will show that, in this case, the wait predicate
$|\{k~:~answers_i[k][i][j]=\bot\}|>\frac{n+t}{2}$  eventually becomes true.
\end{itemize}

We  first show that, if a correct simulator receives
{\sc value'ack}$(j,w)$ from more than $\frac{n+t}{2}$ different simulators,
then all correct simulators do receive {\sc value'ack}$(j,w)$ from more
than $\frac{n+t}{2}$ different simulators.
If a correct simulator receives
{\sc value'ack}$(j,w)$ from more than $\frac{n+t}{2}$ different simulators,
at least $(t+1)$ correct simulators broadcast it.
Every correct simulator will then receive the message {\sc value'ack}$(j,w)$
from at least $(t+1)$ different simulators and, if not already done,
 broadcasts it (lines~B\ref{SA-B-24}-B\ref{SA-B-25}).
All correct simulators will then receive the message {\sc value'ack}$(j,w)$
from at least $n-t>\frac{n+t}{2}$ different simulators.\\

According to the previous observation, let us
consider the case in which no correct simulator ever receives  the message
{\sc value'ack}$(j,w)$  from more than $\frac{n+t}{2}$different simulators.
A correct simulator $q_k$ assigns a non-$\bot$ value to $values_k[j]$ only
if  it receives {\sc value'witness}$(j,w)$ from more than
$\frac{n+t}{2}$ different simulators (line~B\ref{SA-B-22}), or if it receives
{\sc value'ack}$(j,w)$ from more than $\frac{n+t}{2}$ different simulators
(line~B\ref{SA-B-25}). If a correct simulator receives
{\sc value'witness}$(j,w)$ from more than $\frac{n+t}{2}$ different
simulators, according to Theorem~\ref{theo-generic-pattern}, all correct
simulators  receive {\sc value'witness}$(j,w)$ from more than
$\frac{n+t}{2}$ different simulators, and  broadcast
the message {\sc value'ack}$(j,w)$.
Because no correct simulator ever receives {\sc value'ack}$(j,v_j)$ messages
from more than $\frac{n+t}{2}$ different simulators, no correct simulator $q_k$
will ever assign a non-$\bot$ value to $values_k[j]$ (line~B\ref{SA-B-22}).

When a correct simulator receives a {\sc read}$(i,j)$ message from $q_i$, it
waits until it has received {\sc value'ack}$(i,v_i)$ messages
from more than $\frac{n+t}{2}$ different simulators (line~B\ref{SA-B-24}).
The reasoning above (first item)
shows that this will eventually become true.

Every correct simulator $q_k$ will then broadcast
{\sc read'answer}$(i,j,\bot)$. This will cause all correct simulators to
broadcast mess sages {\sc read'answer'witness}$(k,i,j,\bot)$,
which will be received by  the simulator $q_i$. This
will then assign $\bot$ to $answers_i[k][i][j]$ for
at least $n-t > \frac{n+t}{2}$ different values of $k$.
Consequently, it  will not remain blocked at
lines~B\ref{SA-B-05}-B\ref{SA-B-06}.

\vspace{-0.2cm}
\item ${\sf wait}$ instruction at line B\ref{SA-B-13}.\\
As simulator $q_i$  broadcasts its view with a message
{\sc view}$(i,view)$  (line~B\ref{SA-B-12}),
every correct simulator  checks if this view is
consistent when it receives it (lines~B\ref{SA-B-36}-B\ref{SA-B-41}).
Let us first consider the entries $view[j]$ such that $view[j]=w\neq\bot$.
This means that $q_i$ has received {\sc value'ack}$(j,w)$ from more than
$\frac{n+t}{2}$ different simulators. All the
correct simulators then receive the same message from a sufficient number of
different simulators and do not remain blocked at line~B\ref{SA-B-38}
(Theorem~\ref{theo-generic-pattern}).

Let us now consider the entries $view[j]$ such that $view[j]=\bot$.
Simulator $q_i$ assigned $\bot$ to $view[j]$
 because it received
{\sc read'answer'witness}$(k,i,j,\bot)$ from more than
$\frac{n+t}{2}$ different simulators (lines~B\ref{SA-B-32}-\ref{SA-B-33}).
 According to 
Theorem~\ref{theo-generic-pattern}, all the correct simulators $q_x$
will also receive {\sc read'answer'witness}$(k,i,j,\bot)$ from more than
$\frac{n+t}{2}$ different simulators, and will assign $\bot$ to
$answers_x[k][i][j]$. They will thus not remain blocked at line~B\ref{SA-B-39}.

All the correct simulators will then broadcast the message
{\sc view'witness}$(i,view)$ (line~B\ref{SA-B-42}).
By Theorem~\ref{theo-generic-pattern}, they will
all send {\sc view'ack}$(i,view)$ to $q_i$. This will allow $q_i$ to
terminate its invocation of ${\sf propose}(i,v_i)$, which concludes the
proof of the lemma.
\end{itemize}
\vspace{-0.6cm}
\renewcommand{\toto}{lemma-termination-if-valid}
\end{proofL}

\begin{lemma}
\label{lemma-valid-if-correct}
Let $v_1,\ldots,v_x,\ldots$ be the values proposed by a correct simulator
$q_i$ to a sequence of safe agreement objects. If $q_i$ does not invoke
${\sf propose}()$ operations concurrently and ${\sf valid}_j(i,v_x)$ is
eventually satisfied at every correct simulator $q_j$, then
${\sf valid}_j(i,v_{x+1})$ is also eventually satisfied at $q_j$.
\end{lemma}

\begin{proofL}
We  consider here that the sub-predicate $P2$ is always satisfied, and
thus consider only the sub-predicate $P1$. Let us recall that $P1$ states
that, for every message {\sc value}$(i,-)$ that $q_j$ received from $q_i$,
and for every message {\sc value'valid}$(i,-)$ that $q_j$ received from 
at least $t+1$ different simulators, it has also
received the corresponding messages {\sc view'witness}$(i,-)$.

By hypothesis, ${\sf valid}_j(i,v_x)$ is eventually satisfied at the
correct simulator $q_j$. Once $q_i$ broadcasts the message
{\sc value}$(i,v_x)$, $q_j$  only needs  to receive the corresponding
{\sc view'witness}$(i,view)$ for $P1$ to be satisfied.
By Lemma \ref{lemma-termination-if-valid}, $q_i$ terminates its invocation
of ${\sf propose}(i,v_x)$, from which we conclude that it  received
{\sc view'ack}$(i,view)$ from more than  $\frac{n+t}{2}$ different simulators
(line~B\ref{SA-B-13}).
A correct simulator sends such a message only if it has received
{\sc view'witness}$(i,view)$ from more than  $\frac{n+t}{2}$ different
simulators (lines~B\ref{SA-B-47}-B\ref{SA-B-48}).
According to Theorem~\ref{theo-generic-pattern},
all the correct simulators  also broadcast it
(lines~B\ref{SA-B-44}-B\ref{SA-B-45}). The correct simulator $q_j$  then
receives them from more than $\frac{n+t}{2}$ different simulators.
The predicate ${\sf valid}_j(i,v_{x+1})$ is then eventually satisfied at 
$q_j$.
\renewcommand{\toto}{lemma-valid-if-correct}
\end{proofL}

\paragraph{Decide-termination}

\begin{lemma}
\label{lemma:one-decides-all-decide}
If a correct simulator terminates its invocation of ${\sf decide}()$, 
then all correct simulators terminate their invocation of ${\sf decide}()$.
\end{lemma}

\begin{proofL}
Suppose, by way of contradiction, that the invocation of ${\sf decide}()$
by a correct simulator $q_i$ terminates, and that the invocation of 
${\sf decide}()$ by another correct simulator $q_j$ does not.

The invocation of ${\sf decide}()$ by $q_i$ can terminate only if 
the predicate at lines~C\ref{SA-C-15}-C\ref{SA-C-16} is satisfied.
Let $q_k$ be any simulator in the set $\sigma$ defined at 
line~C\ref{SA-C-15}. We  show that $all\_views_i[k] = view$ implies that 
we  eventually have $all\_views_j[k] = view$, and thus that $q_j$ must 
decide.

Simulator $q_i$ assigns $view$ to $all\_views_i[k]$ at line~B\ref{SA-B-48}. 
This can happen only because $q_i$ received 
{\sc view'witness}$(k,view)$ messages from more than 
$\frac{n+t}{2}$ different simulators. According to 
Theorem~\ref{theo-generic-pattern},  $q_j$  eventually  receives
enough {\sc view'witness}$(k,view)$ messages
and  also assigns $view$ to $all\_views_j[k]$. Simulator
$q_j$ will then also have to decide.
\renewcommand{\toto}{lemma:one-decides-all-decide}
\end{proofL}

\begin{lemma}
\label{lemma:at-most-one-blocked-byz}
The invocations of ${\sf  decide}()$ by all the correct simulators 
on all the safe agreement objects terminate, except for at most $t$
safe agreement objects.
\end{lemma}

\begin{proofL}
Suppose, by way of contradiction, that there are $t+1$ safe agreement 
objects such that at least one correct simulator never terminates its 
invocation of ${\sf decide}()$. By Lemma 
\ref{lemma:one-decides-all-decide}, there must be 
 $(t+1)$ different safe agreement objects in which 
no correct simulator  terminates its invocations of ${\sf decide}()$.

The invocation of the ${\sf decide}()$ operation by a correct simulator 
$q_i$ on a safe agreement object can only be blocked at 
lines~C\ref{SA-C-15}-C\ref{SA-C-16}, if the corresponding predicate is 
never satisfied. This can happen if (1) there is no simulator $q_j$ such 
that $all\_views_i[j] \neq \bot$ or, (2) for every non-empty set of 
simulators $\sigma$, there are  two  simulators $q_y\in\sigma$ and  $q_z$ 
such that $all\_views_i[y][z] \neq \bot \wedge all\_views_i[z] = \bot$.
Because a correct simulator $q_i$ invokes  ${\sf propose}()$  
before invoking ${\sf decide}()$, case (1) cannot happen; we always have 
$all\_views_i[i] \neq \bot$. We then consider case (2). 

Case (2) can happen if $q_z$ starts an invocation of 
${\sf propose}()$ and communicates its proposed value to other processes, 
but does not terminate its invocation by communicating its view.
Because there are at most $t$ faulty simulators, by the pigeonhole 
principle, there must be a faulty simulator $q_z$ that prevents $q_i$ from 
deciding on two different safe agreement objects.

A correct simulator $q_k$ broadcasts a {\sc value'valid}$(z,-)$
after receiving a {\sc value}$(z,-)$ message only if the predicate 
${\sf valid}_k(z,-)$ is satisfied (line~B\ref{SA-B-15}). 
Due to the predicate $valid_k(z,-)$, this is true only if $q_k$ received 
{\sc view'witness}$(z,-)$ messages from at least $(n-t)$ different 
simulators, each of these messages corresponding to the all the  
{\sc value}$(z,-)$ and {\sc value'valid}$(z,-)$ messages 
that it has previously received (see the definition of the predicate $P1$ of 
${\sf valid}_k(z,-)$). 

Let ${\sf propose}(v_1)$ be the invocation of ${\sf propose}()$ by $q_z$ on 
the first safe agreement object on which $q_i$ is blocked, and 
${\sf propose}(v_2)$ the one on the second safe agreement object on which 
$q_i$ is blocked. Because there is a simulator $q_y\in\sigma$ such that 
$all\_views_i[y] \neq \bot$ in the two invocations of ${\sf decide}()$ by 
$q_i$, in both cases, more than $\frac{n+t}{2}$ different simulators
have broadcast a {\sc view'witness}$(y,-)$ message (line~B\ref{SA-B-48}). 
Both sets  include correct simulators. They 
must then have received {\sc value'ack}$(z,v_1)$ and 
{\sc value'ack}$(z,v_2)$ from more than $\frac{n+t}{2}$ different 
simulators (line~B\ref{SA-B-38}). Again, both sets  include correct 
simulators that must have received {\sc value'witness}$(z,v_1)$ and 
{\sc value'witness}$(z,v_2)$ from more than $\frac{n+t}{2}$ different 
simulators (line~B\ref{SA-B-21}).

In order to broadcast a {\sc value'witness}$(z,-)$ message,
a correct simulator must either (a) receive {\sc value'witness}$(z,-)$ 
messages from at least $t+1$ different simulators (line~B\ref{SA-B-18}),
or (b) receive {\sc value'valid}$(z,-)$ messages from more than $\frac{n+t}{2}$ 
different simulators (line~B\ref{SA-B-16}). The first correct simulator 
that broadcasts a 
{\sc value'witness}$(z,-)$ message must then have received 
{\sc value'valid}$(z,-)$ messages from more than $\frac{n+t}{2}$ 
different simulators.

According to Lemma 
\ref{lemma:quorum-intersection}, there is a least one correct simulator 
$q_\ell$ that broadcasts both {\sc value'valid}$(z,-)$ messages 
(line~B\ref{SA-B-15}). 
In order to 
do so, the predicate ${\sf valid}_\ell(z,v_2)$ must have been verified 
 at the time that $q_\ell$  broadcast the 
{\sc value'valid}$(z,v_2)$ message. It must then have received 
the {\sc view'witness}$(z,view)$ messages that correspond to $v_1$ 
from more than $\frac{n+t}{2}$ different 
simulators. According to Theorem~\ref{theo-generic-pattern},
$q_i$ must then also have received these messages from more than 
$\frac{n+t}{2}$ different 
simulators and assigned $view$ to $all\_views_i[z]$ (line~B\ref{SA-B-48}) 
in the instance that corresponds to the invocation of ${\sf propose}(v_1)$ 
by $q_z$, a contradiction that concludes the proof of the lemma.
\renewcommand{\toto}{lemma:at-most-one-blocked-byz}
\end{proofL}

\paragraph{Agreement}

\begin{lemma}
For any simulator $q_x$ and any correct simulator $q_i$, if $q_i$ assigns a
non-$\bot$ value $v$ to $values_i[x]$,
then  $(1)$ no  value $v'\neq  v$ is  ever assigned  to $values_j[x]$  by a
correct  simulator $q_j$  and    $(2)$ each  such  correct simulator  $q_j$
eventually assigns $v$ to $values_j[x]$.
\label{lemma-byz-uniform}
\end{lemma}

\begin{proofL}
Let $q_k$ be the first simulator that assigns $v$ to $values_k[x]$.
Since $q_k$ executes line~B\ref{SA-B-22},
it received strictly  more than $\frac{n+t}{2}$ {\sc value'witness}~$(x,v)$
messages from different simulators.
At least  $t+1$ correct  simulators consequently sent  this message  to all
processes at line~B\ref{SA-B-17} or at line~B\ref{SA-B-19}.
By Theorem~\ref{theo-generic-pattern},
every  correct  simulator  $q_j$  consequently eventually  receives  such  a
message from each correct simulator and assigns $v$ to $values_j[x]$.

Suppose that there exists a value $v'\neq v$ such that
there is a correct simulator $q_\ell$ that assigns $v'$ to $values_\ell[x]$.
Suppose that $q_\ell$ is the first process to do so.
It follows that $q_\ell$ received {\sc value'witness}~$(x,v')$ messages
from strictly more than $\frac{n+t}{2}$ different processes 
(line~B\ref{SA-B-21} or line~B\ref{SA-B-24}).

Consider the first correct simulator that broadcasts a 
{\sc value'witness}~$(x,v')$ message. In order to do so, it must have 
received {\sc value'valid}~$(x,v')$ messages
from strictly more than $\frac{n+t}{2}$ different processes 
(lines~B\ref{SA-B-16}-B\ref{SA-B-17}).
However, the first correct simulator that broadcasts a 
{\sc value'witness}~$(x,v)$ message must also have 
received {\sc value'valid}~$(x,v)$ messages
from strictly more than $\frac{n+t}{2}$ different processes.
There must then be a correct simulator that sent both 
{\sc value'valid}~$(x,-)$ messages. The only place a correct simulator can 
send a {\sc value'valid}~$(x,-)$ message is at Line~\ref{SA-B-15} and it 
does so only once for each simulator $q_x$, a contradiction which concludes 
the proof of the lemma.
\renewcommand{\toto}{lemma-byz-uniform}
\end{proofL}

\begin{lemma}
For any  simulators $q_k, q_\ell, q_x$  and any correct  simulator $q_i$, if
$q_i$ assigns a non-$\bot$ value $v$  to $answers_i[\ell][k][x]$, then
$(1)$ no value $v'\neq v$ is ever assigned to $answers_j[\ell][k][x]$ by
a  correct simulator  $q_j$ and  $(2)$ each  such  correct  simulator $q_j$
eventually assigns $v$ to $answers_j[\ell][k][x]$.
\label{lemma-byz-uniform-reads}
\end{lemma}

\begin{proofL}
The proof is the same as for Lemma~\ref{lemma-byz-uniform}.
\renewcommand{\toto}{lemma-byz-uniform-reads}
\end{proofL}

\begin{lemma}
For any simulator $q_x$ and any correct simulator $q_i$, if $q_i$ assigns a
non-$\bot$ value $view$   to $all\_views_i[x]$,
then (1) no value $view'\neq view$ is ever assigned to $all\_views_j[x]$ by
a  correct simulator  $q_j$  and   (2) each  such  correct simulator  $q_j$
eventually assigns $view$ to $all\_views_j[x]$.
\label{lemma-byz-uniform-view}
\end{lemma}

\begin{proofL}
The proof is the same as for Lemma~\ref{lemma-byz-uniform}.
\renewcommand{\toto}{lemma-byz-uniform-view}
\end{proofL}

\begin{lemma}
\label{lemma:safe-byz-agreement}
No two invocations of ${\sf decide}()$ return different values.
\end{lemma}

\begin{proofL}
Let us recall that the algorithm implementing the operation 
$\mathsf{decide}()$
is described at lines~C\ref{SA-C-15}-C\ref{SA-C-19}.
Let $q_i$ and $q_j$ be two correct simulators.
According  to  Lemmas~\ref{lemma-byz-uniform}-\ref{lemma-byz-uniform-view},
we have:
\begin{itemize}
\vspace{-0.2cm}
\item $(values_i[x]\neq \bot) ~\wedge~(values_j[x]\neq \bot)
      ~\Rightarrow~(values_i[x]=values_j[x])$.
\vspace{-0.2cm} 
\item $(answers_i[\ell][k][x]\neq?\land answers_j[\ell][k][x]\neq?)
      \Rightarrow(answers_i[\ell][k][x]=answers_j[\ell][k][x])$.
\vspace{-0.2cm} 
\item $(all\_views_i[x]\neq \bot ~\wedge~ all\_view_j[x]\neq \bot)
      ~\Rightarrow~(all\_views_i[x]=all\_view_j[x])$.
\end{itemize}  
 
Let us assume, by contradiction, that $q_i$ and $q_j$
decide different values. This means that the sets  $min\_\sigma_i$ and
 $min\_\sigma_j$ computed at line~C\ref{SA-C-17} by   $q_i$  and  $q_j$,
respectively,  are different.

Since $min\_\sigma_i$ and $min\_\sigma_j$ are different,
let us consider $z\in min\_\sigma_i\setminus min\_\sigma_j$
(if $min\_\sigma_i \subsetneq min\_\sigma_j$,  swap $i$ and $j$).
According to the closure predicate used at line~C\ref{SA-C-16},
as $z\notin min\_\sigma_j$,
we have  $\forall y\in min\_\sigma_j~:~all\_views_j[y][z]=\bot$.

It  follows  that $q_j$  received  {\sc view'witness}  ($y,all\_view_j[y]$)
messages  (with $all\_view_j[y][z]=\bot$)
from  a  set  of  simulators   $Q_{j,vw}$  of  size  strictly  larger  than
$\frac{n+t}{2}$ (the subscript $vw$ stands for ``view witness'').
The correct simulators  of $Q_{j,vw}$ sent these messages  after checking at
line~B\ref{SA-B-39} that
a set $Q_{j,vw,r}$ of strictly more than $\frac{n+t}{2}$ reliably broadcast
(thanks to the mechanism of lines~B\ref{SA-B-26} to B\ref{SA-B-33})
a {\sc read'answer} ($y,z,\bot$) message.
The   correct   simulators  of   $Q_{j,vw,r}$   sent   these  messages   at
line~B\ref{SA-B-26} after they  received  {\sc value'ack} ($y,v_y$) messages
from a set $Q_{y,w}$ of strictly more than $\frac{n+t}{2}$ simulators
 (the subscript $w$ stands for ``witness'').
Each  correct simulator $q_k$  of $Q_{y,w}$  had $values_k[y]=v_y$  when it
sent this message and it happens
strictly  before the  first  correct simulator  sends  a {\sc  read'answer}
($y,z,\bot$) message.

Since $z\in min\_\sigma_i$, the correct simulator $q_i$ received
{\sc view'witness} ($z,all\_view_i[z]$) messages
from a set $Q_{i,vw}$ of strictly more than $\frac{n+t}{2}$ simulators.
The correct simulators of $Q_{i,vw}$ sent these messages
after the check of the values at lines~B\ref{SA-B-38}-B\ref{SA-B-39}.

Suppose that some of them verified the predicate of line~B\ref{SA-B-39}
for $x=y$.
It entails  that a set  $Q_{i,vw,r}$ of strictly more  than $\frac{n+t}{2}$
simulators reliably broadcast a {\sc read'answer} ($z,y,\bot$).
The correct simulators of $Q_{i,vw,r}$ sent this message after receiving at
line~B\ref{SA-B-24}  {\sc value'ack} ($z,v_z$) messages from a set $Q_{z,w}$
of strictly more than $\frac{n+t}{2}$ simulators.
This  happens  strictly before  the  first  {\sc read'answer}  ($z,y,\bot$)
message is sent by a correct simulator.
Since       $|Q_{i,vw,r}|,|Q_{j,vw,r}|>\frac{n+t}{2}$,      $Q_{i,vw,r}\cap
Q_{j,vw,r}$  contains
at least a correct simulator $p_k$.

Simulator $p_k$ thus broadcast a {\sc read'answer} ($y,z,\bot$) message
and a {\sc read'answer} ($z,y,\bot$) message (line B\ref{SA-B-26}). It then 
had  $views_k[z] = \bot$ before broadcasting the  
{\sc read'answer} ($y,z,\bot$) message and $views_k[y] = \bot$ before 
broadcasting the {\sc read'answer} ($z,y,\bot$).
Because of the first instruction of line~B\ref{SA-B-25} this is impossible,
and thus each correct
process that sends a {\sc view'witness} ($z,all\_views_i[z]$) message ended
the wait instruction
of   lines~B\ref{SA-B-38}-B\ref{SA-B-39}  by   verifying  the   predicate  of
line~B\ref{SA-B-38}. 
This entails that
$\forall x\in\Pi~:~all\_views_x[z]\neq\bot \Rightarrow
all\_views_x[z][y]\neq\bot$.   Consequently, $all\_views_i[z][y]\neq\bot$.

Since $z\in min\_\sigma_i$, $all\_views_i[z]\neq\bot$ and thus
$all\_views_i[z][y]\neq\bot$.
According to the predicate of line~C\ref{SA-C-16}, this entails that
$y\in min\_\sigma_i$, and since the previous reasoning holds for any
$y\in min\_\sigma_j$, it shows that $min\_\sigma_j\subseteq min\_\sigma_i$.
It follows that, when $q_i$ executes line~C\ref{SA-C-17}, $\forall y\in
min\_\sigma_j~:~all\_views_i[y]\neq\bot$ and,
consequently, $\forall y\in
min\_\sigma_j~:~all\_views_i[y]=all\_views_j[y]$. It entails that if
$|min\_\sigma_j|<|min\_\sigma_i|$,
then $min\_\sigma_j$ would have been chosen by $q_i$ at
line~C\ref{SA-C-17}, which proves that $min\_\sigma_i=min\_\sigma_j$
and contradicts the fact that $q_i$ and $q_j$ decide differently.
\renewcommand{\toto}{lemma:safe-byz-agreement}
\end{proofL}

\paragraph{Correct values are valid}
\begin{lemma}
\label{lemma-correct-is-valid}
If a correct simulator $q_i$ decides the value $v$, there is 
a correct simulator $q_j$ such that ${\sf valid}_j(-,v)$.
\end{lemma}
\begin{proofL}

Let $v$ be the value decided by a correct simulator $q_i$. Value $v$ has
then be proposed by a simulator $q_j$ such that $all\_views_i[j]\neq \bot$
(definition of $\sigma$ at lines~\ref{SA-C-15}-C\ref{SA-C-16} 
and choice of value at 
line~C\ref{SA-C-18}). In order to assign a non-$\bot$ value to 
$all\_views_i[j]$, $q_i$ must have received {\sc view'witness}$(j,-)$
messages from more than $\frac{n+t}{2}$ different  simulators 
(lines~B\ref{SA-B-47}-B\ref{SA-B-48}), and 
consequently from at least one correct 
simulator. Consider the first correct simulator $q_x$ that has broadcast a
{\sc view'witness}$(j,-)$ message. Before sending it, it must have assigned 
a non-$\bot$ value to $values_x[j]$ (lines B\ref{SA-B-35}-B\ref{SA-B-42}).
It then has received either (a) {\sc value'witness}$(j,-)$ messages from 
more than $\frac{n+t}{2}$ different  simulators or 
(b) {\sc value'ack}$(j,-)$ messages from more than $\frac{n+t}{2}$ different  
simulators.

In case (a), consider the first correct simulator $q_k$ that has 
broadcast a {\sc value'witness}$(j,-)$ message. In order to do so, 
it must have received {\sc value'valid}$(j,-)$ messages from more 
than $\frac{n+t}{2}$ different  
simulators (lines~B\ref{SA-B-16}-\ref{SA-B-17}).
The predicate ${\sf valid}_k(j,v)$ must have been satisfied at the 
simulators that broadcast these messages(line~B\ref{SA-B-15}). 
In case (b), the first correct simulator  that has 
broadcast a {\sc value'ack}$(j,-)$ message must first have received 
{\sc value'witness}$(j,-)$ messages from more than $\frac{n+t}{2}$ 
different  simulators (lines B\ref{SA-B-21}-B\ref{SA-B-23}). 
The situation is then similar to Case (a).
\renewcommand{\toto}{lemma-correct-is-valid}
\end{proofL}

\begin{theorem}
\label{theorem-SA-object-is-correct}
The algorithms described in 
Figure~{\em\ref{algo-SA-msg-passing-byzantine-propose}} and 
 Figure~{\em\ref{algo-SA-msg-passing-byzantine-answers}} implement a 
safe-agreement object in  $\BM[t<n/3]$.
\end{theorem}

\begin{proofT}
The proof follows from the previous lemmas. 
\renewcommand{\toto}{theorem-SA-object-is-correct}
\end{proofT}

\subsection{Simulation algorithm and its proof in  $\BM[t<n/3]$}
\label{sec:simulation-byzantine}
\paragraph{Simulation algorithm}
When we consider the simulation algorithm described in
Figure~\ref{algo:simulation-crash}, we observe that the $n$ simulators
communicate  only through  safe agreement objects.  It follows  that
the same algorithm works in $\BM[t<n/3]$, when the crash-tolerant
safe agreement objects are replaced by Byzantine-tolerant
safe agreement objects previously described.
Two  things remain to be done: define the specific sub-predicate
$P2$ of the predicate ${\sf valid}()$, and
do a specific proof of this algorithm
(i.e., a proof based on the specification of the  Byzantine-tolerant
safe agreement objects defined in Section~\ref{sec:spec-Byzantine-SA}).

\paragraph{Sub-predicate $P2$}
As far as $P2$ is concerned we have the following.
Let us consider the simulator $q_i$ that invokes  ${\sf valid}_i(j,v)$,
with respect to the simulation of a process $p_x$.
In the simulation algorithm, the parameter $v$ is the message $msg$
that $q_j$ proposes to a safe agreement object from which will be decided
the next message to be received  by the simulated process $p_x$
(lines~\ref{Simu-C-08}-\ref{Simu-C-09} of Figure~\ref{algo:simulation-crash}).
$P2$ checks, from $q_i$'s local point of view,  that, if the message $v$
has been sent in the simulation, then it has not yet been consumed,
i.e., $(v\in sent_i[x])~\Rightarrow~(v \notin received_i[x])$.

\paragraph{Proof of the simulation algorithm in $\BM[t<n/3]$}

\begin{lemma}
\label{lemma:at-most-one-crash-byz}
The simulation of at most $t$ simulated processes can be blocked.
\end{lemma}

\begin{proofL}
The only places where a correct simulator $q_i$ can block is 
during the invocation of
the safe agreement operation ${\sf decide}()$. Such invocations appear at
line~\ref{Simu-C-02}, and line~\ref{Simu-C-11}.

Because the invocations by all the correct simulators  of ${\sf  decide}()$
on all the safe agreement objects terminate, except for at most $t$
safe agreement objects (Lemma \ref{lemma:at-most-one-blocked-byz}),
the simulation of at most $t$ simulated processes can be blocked.
\renewcommand{\toto}{lemma:at-most-one-crash-byz}
\end{proofL}

\begin{lemma}
\label{lemma:same-seq-of-messages-byz}
The simulation of the reception of the $k$-th
message received by a simulated process $p_j$, returns the same message
at all correct simulators.
\end{lemma}

\begin{proofL}
The simulation of the message receptions for a simulated process $p_j$,
are executed at each correct simulator $q_i$ at lines~\ref{Simu-C-08}-\ref{Simu-C-11},
and all the correct simulators use the same sequence of sequence numbers
(line~\ref{Simu-C-07}).
It then follows from the agreement property of the safe agreement object
$\SA[j,sn]$, that no two correct simulators obtain different messages
when they invoke  $\SA[j,sn].{\sf decide}()$, and the lemma follows.
\renewcommand{\toto}{lemma:same-seq-of-messages-byz}
\end{proofL}

\begin{lemma}
\label{lemma:one-dec-value-byz}
For every simulated processes $p_j$, no two correct simulators return
different values.
\end{lemma}

\begin{proofL}
The only non-deterministic elements of the simulation are the input vectors
$input_i[1..n']$ at each simulator $q_i$, and the reception of the simulated
messages.

The lines~\ref{Simu-C-01}-\ref{Simu-C-02} of the
simulation force  the correct simulators to agree on the same input value
for each simulated process $p_j$, $1\leq j \leq n'$.
Similarly,  as shown by Lemma~\ref{lemma:same-seq-of-messages-byz},
for each simulated process $p_j$, the lines~\ref{Simu-C-07}-\ref{Simu-C-11}
direct the simulators to  agree on the very same sequence of messages
received by $p_j$. It follows from the fact that the function $\delta_j()$
is deterministic, that   any two correct simulators $q_i$ and $q_k$,
that execute  lines~\ref{Simu-C-15}-\ref{Simu-C-16} during the
same ``round number'' $sn_i[j]=sn_k[j]$, are such that $state_i[j]=state_k[j]$,
from which the lemma follows.
\renewcommand{\toto}{lemma:one-dec-value-byz}
\end{proofL}

\begin{lemma}
\label{lemma:correct-simulation-byz}
The sequences of message receptions simulated by each simulator $q_i$
on behalf of each simulated process $p_j$, define a correct execution of
the simulated algorithm.
\end{lemma}

\begin{proofL}
To prove the correctness of the simulation, we have  to show that
\begin{enumerate}
\vspace{-0.2cm}
\item
\label{item0-byz}
Every message that was received by a simulated process was sent
 by another simulated process,
\vspace{-0.2cm}
\item
\label{item1-byz}
Every message that was sent by a simulated process to another simulated
process (whose simulation is not blocked either), is received, and
\vspace{-0.2cm}
\item
\label{item2-byz}
The simulated messages respect a simulated physical order
(i.e., no message is ``received'' before being  ``sent'').
\end{enumerate}

Item~\ref{item0-byz} follows from Lemma \ref{lemma-correct-is-valid}
 and from the definition of $P2$.
Item~\ref{item1-byz} is satisfied because the messages sent by
the simulated process $p_j$ to the simulated process $p_k$ are received
(lines~\ref{Simu-C-09}-\ref{Simu-C-11}) in their sending order
(as defined at line~\ref{Simu-C-04}  and line~\ref{Simu-C-14}).
Hence, if $p_k$ is not blocked (due to a faulty simulator)
it obtains the messages from $p_j$ in their sending order.

For Item~\ref{item2-byz}, let us define a (simulated) physical order as follows.
For each simulated message $m$, let us consider the first time at which the
reception of $m$  was simulated (i.e., this occurs when
 --for the first time-- a simulator terminates  the invocation  of
$\SA[-,-].{\sf decide}()$  that returns $m$).
A message that is decided has been proposed by a simulator to a safe agreement
object before  being decided (validity property).The sending time  of a
simulated message is then the first time at which
$\SA[-,-].{\sf propose}(m)$ is  invoked by a simulator.
It follows that any simulated  message is sent before being received,
which concludes the lemma.
\renewcommand{\toto}{lemma:correct-simulation-byz}
\end{proofL}

\begin{lemma}
\label{lemma:nb-simulated-proc-byz}
Each correct simulator $q_i$ computes the decision value of
at least $(n'-t)$ simulated processes.
\end{lemma}

\begin{proofL}
Due to Lemma~\ref{lemma:at-most-one-crash-byz}, and the fact that at
most $t$ simulators may be byzantine, it follows that at most $t$
simulated processes may be prevented from progressing. As (a) by
assumption the simulated algorithm $A'$ is $t$-resilient, and (b) due
to Lemma~\ref{lemma:correct-simulation-byz} the simulation produces a
correct simulation of $A'$, it follows that at least $(n'-t)$
simulated processes decide a value.
\renewcommand{\toto}{lemma:nb-simulated-proc-byz}
\end{proofL}

\begin{theorem}
\label{theorem:main-simulation-byz}
Let $A$ be an algorithm solving a decision task in $\CMprim[t<n']$.
The algorithm described in Figure~\ref{algo:simulation-crash},
in which Byzantine-tolerant safe agreement objects are used, is a
correct simulation of $A$ in~$\BM[t<n/3]$.
\end{theorem}

\begin{proofT}
The theorem follows from Lemma~\ref{lemma:correct-simulation-byz} and
Lemma~\ref{lemma:nb-simulated-proc-byz}.
\renewcommand{\toto}{theorem:main-simulation-byz}
\end{proofT}

Additionally, the reader can easily check that the simulation of a message only requires 
a polynomial number of messages in the base system, and the increase in size of these messages,
when compared to the size of the simulated message, is
also polynomial.

\section{Implications of the Simulation} \label{sec:conclusion}
\paragraph{BG-simulation in Byzantine message-passing systems} 
A main result of this  paper is a signature-free
distributed algorithm that solves  
BG-simulation in Byzantine asynchronous message-passing systems.
In addition to being the first  algorithm  that solves
BG-simulation in such a severe failure context, the proposed simulation
algorithm has noteworthy applications as shown below.

\paragraph{From Byzantine-failures to crash 
failures in message-passing systems}
The simulation presented here allows the execution of a 
$t$-resilient crash-tolerant algorithm in an asynchronous
message-passing system where up to $t$ processes may be Byzantine.
A  feature that is sometimes required from a Byzantine-tolerant algorithm
solving a task (not usually  considered in the crash failure case)
is that the value decided by any correct process should be based only
on  inputs of  correct processes. This prevents Byzantine processes
from ``polluting'' the computation with their inputs.
 A way to guarantee that an input has
been proposed by a correct process is to check that it has been
proposed by at least $(t+1)$ different processes. Assuming that
in any execution at most $m$ values are proposed,  this constraint
translates as $n-t>mt$~\cite{HKR14,MTH14}.

In the case of the simulation presented in
Section~\ref{sec:BG-byzantine-model}, this requirement can easily be
satisfied by adding a first step of computation before the start of
the simulation. Simulators first broadcast their input. They then echo
every value that they receive from more than $t+1$ different
simulators, and consider these values (and only these values) as valid
inputs. An input considered valid by a correct simulator is then
eventually considered valid by all correct simulators, and the only
inputs allowed in the simulation are inputs of correct simulators.
Because we consider colorless tasks, the choice of output is done in the 
same way as in the original BG-simulation: a simulator can adopt the 
output of any simulated process that has decided a value.

The possible Byzantine behaviors are restrained by the underlying
Byzantine-tolerant safe agreement objects used in the simulation.
Surprisingly, this shows that, from the point of view of the
computability of colorless tasks and assuming $n>(m+1)t$ (this
requirement always implies $n>3t$ when at least two different values
can be proposed), Byzantine failures are equivalent to crash-failures.
This provides us with a new understanding of Byzantine failures and
shows that their impact can be restricted to the much simpler
crash-failure case.

\paragraph{From wait-free shared memory to message-passing}
The proposed simulation can be combined with previous works
to further extend the  scope of the result. 
Consider an algorithm $A_0$ that solves a colorless task, where $m>1$,
in a wait-free read/write memory system of $t+1$ processes, denoted 
${\cal CARW}_{t+1,t}[\emptyset]$.
Using the basic BG-simulation~\cite{BG93}, 
this algorithm can be transformed into an algorithm $A_1$ 
that works in the $t$-resilient read/write memory system 
of $(m+1)t+1$ processes, in which at most $t$ can crash. 
This model is denoted ${\cal CARW}_{(m+1)t+1,t}[\emptyset]$.
Using an implementation
of a read/write memory in a crash-prone message-passing system in which a 
majority of processes are correct~\cite{ABD95}, we obtain an algorithm  
$A_2$  which work in ${\cal CAMP}_{(m+1)t+1,t}[\emptyset]$
(message-passing system 
system of $(m+1)t+1$ processes, in which at most $t$ can crash; 
notice that $m > 0 \Rightarrow (m+1)t+1>2t)$). 
Finally, using the simulation presented in this paper, we  obtain 
Byzantine-tolerant algorithm $A_3$ which works in 
${\cal BAMP}_{(m+1)t+1,t}[\emptyset]$  (message-passing  system of $(m+1)t+1$ 
processes, of which at most $t$ can be Byzantine; notice that
$m>1 \Rightarrow (m+1)t+1>3t$).

These transformations show that, as far as the computability of
colorless tasks that admit up to $m>1$ different input values is
concerned, an $n$-process Byzantine-prone message-passing system, in
which up to $t < n/(m+1)$ processes can be Byzantine, is equivalent to
a wait-free shared memory system of $t+1$ processes, which at most
commit crash failures.  When considering colorless tasks with $m>1$, a
figure relating these transformations is depicted in
Figure~\ref{figure-stacking}.  Differently from the 
full-information algorithm presented in~\cite{MTH14}, 
the simulation presented in the present paper
(along with~\cite{BG93} and~\cite{ABD95}) allows a {\it direct}
transformation of any wait-free shared-memory algorithm that solves a
colorless task into a message-passing Byzantine-tolerant algorithm.

\label{figure-stacking}
\begin{figure}[h]
\centering{
\hspace{-1.5cm}
\ifpdf
\scalebox{0.6}{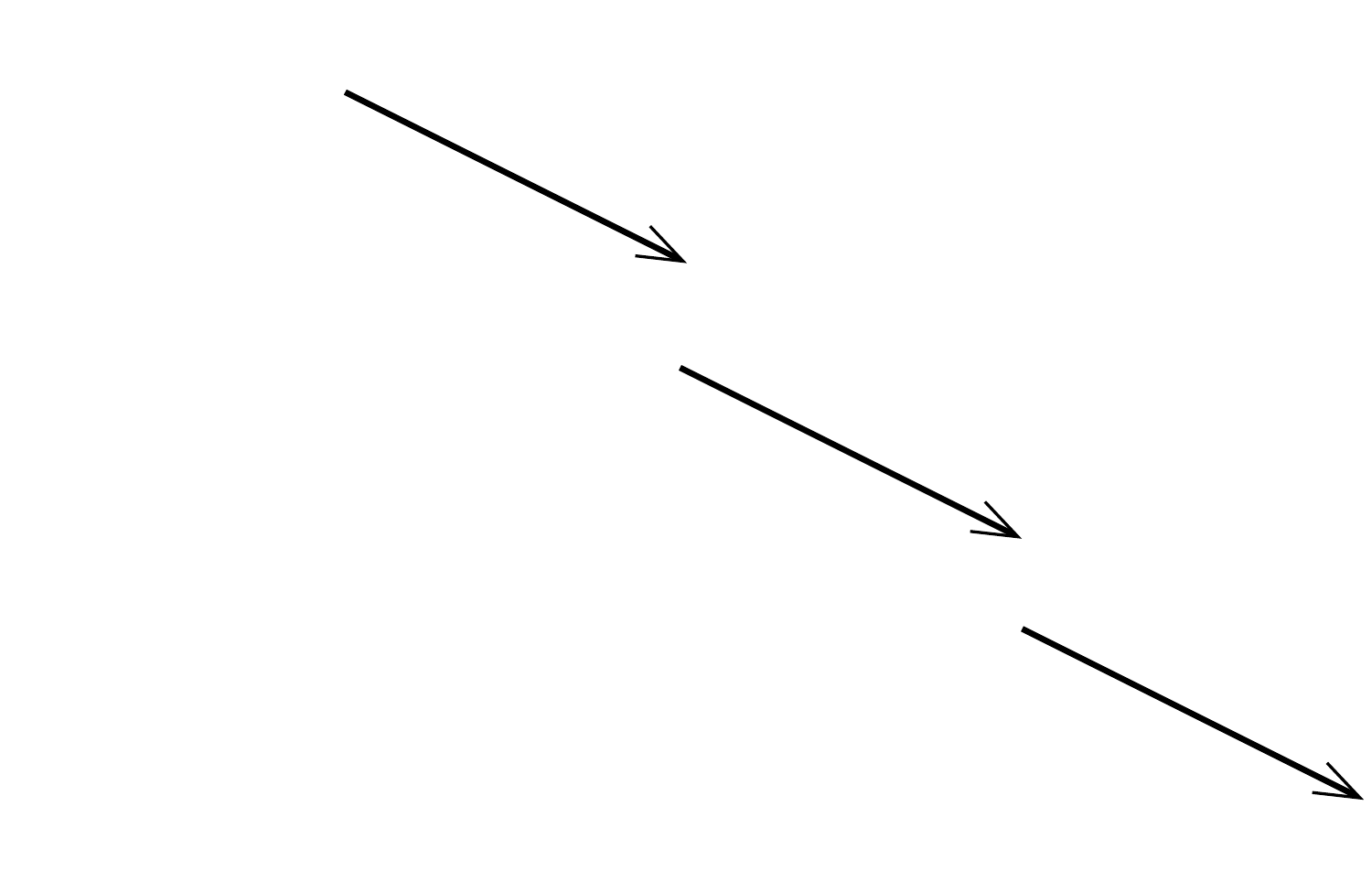}
\else
\scalebox{0.6}{\input{figure-modular.pstex_t}}
\fi
\caption{From crash in read/write to Byzantine in message-passing (with $m>1$)}
\label{modular-construction}
}
\end{figure}

\section*{Acknowledgments}
This work has been partially supported by the Franco-German DFG-ANR Project 
40300781. This project (named DISCMAT) is devoted to mathematical methods in 
distributed computing.
The authors would like to thank Sergio Rajsbaum for discussions on the
BG simulation.

\end{document}